\documentclass[aps,prd,%
tightenlines,%
preprint,draft,%
eqsecnum,%
onecolumn,%
showpacs,%
floats,%
nofootinbib,%
amsfonts,amssymb,amsmath%
]{revtex4}

\usepackage[final]{graphicx}  
\newcommand{\Melt}[3]{\mbox{$\langle {#1} \| {#2} \| {#3} \rangle$}}

\def\Branch{{\mathcal B}}
\def\Fall{{\mathcal F}}
\def\Gauge{{\mathcal G}}
\def\scri{{\mathcal I}}

\begin{document}

\preprint{gr-qc/0309117}

\title{Generalized Quantum Theory\\ 
of Recollapsing Homogeneous Cosmologies}

\author{David Craig}\thanks{E-mail: dcraig@hamilton.edu}
\affiliation{
Department of Physics, Hamilton College,\\
Clinton, NY 13323}

\author{James B. Hartle}\thanks{E-mail: hartle@physics.ucsb.edu}
\affiliation{
Department of Physics, University of California\\
Santa Barbara, CA 93106-9530}

\date{\today}

\begin{abstract}
A sum-over-histories generalized quantum theory is developed for homogeneous
minisuperspace type A Bianchi cosmological models, focussing on the particular
example of the classically recollapsing Bianchi IX universe.  The decoherence
functional for such universes is exhibited.  We show how the probabilities of
decoherent sets of alternative, coarse-grained histories of these model
universes can be calculated.  We consider in particular the probabilities for
classical evolution defined by a suitable coarse-graining.  For a restricted
class of initial conditions and coarse grainings we exhibit the approximate
decoherence of alternative histories in which the universe behaves classically
and those in which it does not.  For these situations we show that the
probability is near unity for the universe to recontract classically if it
expands classically.  We also determine the relative probabilities of
quasi-classical trajectories for initial states of WKB form, recovering for
such states a precise form of the familiar heuristic ``$J\cdot d\Sigma$'' rule
of quantum cosmology, as well as a generalization of this rule to generic
initial states.
\end{abstract}

%

\maketitle


\section{Introduction}
\label{sec:intro}

Hamiltonian quantum mechanics with its requisite preferred time(s) may need to
be generalized to provide a quantum framework for cosmology where spacetime
geometry fluctuates quantum mechanically and does not specify a fixed notion of
time.  One approach to such a generalization is the sum-over-histories
generalized quantum theory of spacetime geometry, described most completely in
\cite{lesH} where references to the earlier literature may be found.  This is a
formulation of quantum theory in fully four-dimensional spacetime form.  The
essential elements of this sum-over histories formulation are:

\begin{enumerate}

\item {\em Fine-Grained Histories} These are the elements of the set of
four-dimensional histories of spacetime metric and matter field
configurations.  They are the most refined descriptions of the universe
it is possible to give.

\item {\em Coarse-Grained Histories} Partitions of the fine-grained histories
into four-dimensional diffeomorphism invariant classes are called
coarse-grained histories.  Such sets of coarse-grained histories are the most
general notion of alternative describable in spacetime terms 
for which quantum theory predicts probabilities.

\item {\em Decoherence Functional} This is a measure of the quantum mechanical
interference between members of a set of alternative coarse-grained histories. 
It is constructed according to sum-over-histories principles and incorporates
a theory of the universe's initial and final condition.  The decoherence
functional is a natural generalization to closed quantum systems of the
algebraic notion of quantum state \cite{IL94,dac97}.  Sets of histories with
negligible interference between all pairs of members, as measured by the
decoherence functional, are said to decohere, or to be consistent.  It is
logically consistent to assign probabilities in an exhaustive set of
alternative histories when, and only when, that set is decoherent.  It is the
criterion of decoherence, rather than any notion of ``measurement'', which
determines the consistency of the quantitative predictions of the theory. 

\end{enumerate}

This paper applies sum-over-histories generalized quantum theory to a
class of homogeneous minisuperspace cosmological models.%
\footnote{There is a substantial literature on these models.  They were
introduced into quantum cosmology by Misner \cite{Misner69a}.  See MacCallum
\cite{Mac79}, Halliwell \cite{Hal91}, and Wainwright and Ellis \cite{WE97} for
partial guides to the literature.}  %
Other formulations of the quantum mechanics of homogeneous cosmological models
have been proposed by Ashtekar, Tate, and Ugla \cite{ATU93},
by Wald \cite{Wald93} (see further \cite{HW95}), and by Marolf \cite{Marolf95}. 
An implementation of generalized quantum theory for these minisuperspace
models is therefore useful for comparison with these other approaches.  That
is especially the case since the formulation of Wald makes the surprising
prediction that a quantum universe whose expansion is accurately predicted by
classical Einstein dynamics cannot have a nearly classical recontracting
phase.  (Instead, time appears to ``freeze'' as the universe approaches its
classical maximum size.)  In this paper we will show that, in suitable
circumstances, sum-over-histories generalized quantum theory predicts that a
universe may remain classical in both its expanding and recontracting phases.

Section \ref{sec:models}\ revisits the classical homogeneous cosmological
models as an introduction to their later quantization.  Section
\ref{sec:IXLambdaeq0}\ sets out the sum-over-histories generalized quantum
mechanics for the classical models of Section \ref{sec:models}\ for the
especially simple case of a closed Bianchi IX cosmology with a single
homogeneous scalar field and vanishing cosmological constant.  %
(Current observations suggest that $\Lambda$ is small but not zero.  Assuming
$\Lambda$ vanishes, while evidently not realistic, simplifies some 
elements of the analysis by allowing quantum evolution to be restricted to
universes which always recontract.  While the quantization framework described
in Section \ref{sec:IXLambdaeq0}\ is in principle applicable to all type A
Bianchi cosmologies, the case where the universe is allowed to expand forever
-- either classically or quantum mechanically -- will not be considered here.)
Section \ref{sec:dffind} explicitly constructs the decoherence functional for
these models for a specific class of boundary conditions: a ``pure'' initial
state and ``indifferent'' final conditions.  In Section \ref{sec:semiclassical}
we examine its semiclassical predictions for initial conditions that correspond
to a single classical trajectory and show how classical evolution can be an
approximation to quantum mechanical evolution in a universe with expanding and
contracting phases, as well as study more general choices of initial state.

Generalized quantum theory \cite{Har91a,Ish94} is a broad framework for
describing and comparing different formulations of quantum mechanics.  A
reduction to essentials of the general principles of the quantum mechanics of
closed systems \cite{Gri84,Omnsum,GH90a}, the framework provides a natural
language with which to frame questions in cosmology concerning whether
probabilities are consistently assigned by quantum theory to a set of
alternative histories of the universe.  The specific sum-over-histories
implementation of its principles sketched above is but one of several
approaches to a conceptually coherent and manageable quantum theory of
spacetime; 
for lucid reviews of some of them and the difficulties they encounter 
see \cite{KucIsham}.  

\section{Homogeneous Cosmological Models}
\label{sec:models}

In this section we review the essential features of homogeneous
cosmological models necessary for the subsequent discussion of their
quantization.\footnote{A classical general reference is \cite{RS75}. 
Extended discussions can be found in \cite{Mac73,Mac79} and in 
Chap.~7 of \cite{Wald84}.}  A
spatially homogeneous cosmological geometry is a spacetime possessing a
group of isometries whose orbits are a family of spacelike surfaces that
foliate the spacetime \cite{Wald84}.  Using a coordinate $t$ that
labels these spacelike surfaces, the metric of a spatially homogeneous
spacetime may be put in the standard form \cite{Wald84,Misner67-8}
\begin{equation}
ds^2=-L^2(t)\,dt^2 + e^{2\alpha(t)} \left(e^{2\beta(t)}\right)_{ij}
\sigma^i\sigma^j\ .
\label{eq:twoone}
\end{equation}
Here the $\sigma^i$ are $t$-independent spatial one-forms preserved 
by the isometries whose dual vector fields $\sigma_i$ obey
\begin{equation}
\left[\sigma_i, \sigma_j\right] = c^k_{ij} \sigma_k,
\label{eq:twotwo}
\end{equation}
where the $c^k_{ij}$    
are the components of the structure tensor of the Lie algebra of the isometry
group in the $\sigma$ basis.%
\footnote{See Wald \cite[section 7.2]{Wald84} for example.  To avoid possible
confusion, note that while Wald's $\sigma$'s coincide with the $\omega$'s of
MacCallum \cite{Mac79}, MacCallum's structure tensor $c$ is defined with the
sign opposite to that of Wald.   
} %
The quantities $L(t)$ and $\alpha(t)$ are functions of $t$ alone; $\beta(t)$
is a $3\times 3$ traceless symmetric matrix that measures the deviations from
isotropy.  The coordinate volume element of the spatial slice scales like
$\sqrt{h} = \exp(3\alpha(t))$ and the same is true for the overall volume
$(4\pi)^2 \exp(3\alpha(t))$ if the spatial slices are closed.

The possible spatial isometry groups may be classified by their Lie algebras
({\it e.g.} as in \cite{TaubEM}) 
and are usually referred to
in the numbering scheme introduced by Bianchi \cite{Bianchi1897} as ``Bianchi
I'' through ``Bianchi IX'' models.  The (simply-connected covering) group
manifold with its natural metric is the manifold for spatial geometry.  In the
Bianchi I models, for example, the group is generated by the translations of
three-dimensional flat space and the manifold is ${\bf R}^3$.  In the Bianchi
IX models the group is $SU(2)$, $c^k_{ij} = \varepsilon_{ijk}$ (in appropriate
coordinates), and the manifold is the three-sphere $S^3$.  The classical
Friedmann-Robertson-Walker (FRW) models are the most familiar examples of
Bianchi universes: the open FRW universe is of Bianchi type I, the flat
universe type V, and the closed FRW universe is of type IX.

The ``type A" Bianchi models are those for which $c^i_{ij} =0$; the rest are
called ``type B".  We restrict attention to type A models because the action
principle for the type B Bianchi models deduced from that for general
relativity by substitution of the homogeneity ansatz (\ref{eq:twoone}) does
not lead to the correct equations of motion \cite{notypeB}; in these cases the
homogeneity of the spatial metric obstructs the elimination of boundary terms
proportional to the trace of the structure tensor.  The Bianchi types I, II,
${\rm VI}_0, {\rm VII}_0,$ VIII, and IX are all type A.

A variety of matter contents are consistent with homogeneity.  As an
illustrative example we shall restrict attention to a single, minimally
coupled, homogeneous scalar field $\phi(t)$ together with a positive
cosmological constant $\Lambda$. For the action of the scalar field we
take
\begin{equation}
  S_M[g, \phi] = - \frac{1}{2}
    \int d^4 x \sqrt{-g}\ \left[(\nabla\phi)^2 + V_\phi (\phi)\right]
\label{eq:twothree}
\end{equation}
for some potential $V_{\phi}(\phi)$. [We follow, as far as possible, the
conventions of \cite{Wald84} with respect to signature $(-,+,+,+)$, 
definitions of the curvature tensors, the extrinsic curvature of a 
hypersurface, {\it etc.}.  We employ units where $\hbar=c=G=1$.]

A canonical action for the type A Bianchi cosmologies with scalar
matter may be arrived at in the following manner.  First note that for
all the type A models with ``diagonal" matter ($T_{0i} = 0$), as in
our example,   
it is possible classically to choose the one-forms $\sigma^i$ in
(\ref{eq:twoone}) so that the matrix $\beta_{ij}$ is diagonal, and the
$c^k_{ij}$ take their canonical values \cite{RS75,Mac73}; the
Einstein equations guarantee that $\beta$ remains diagonal as time
passes.  It is then traditional to parametrize $\beta$ as
$\beta = {\rm diag}\left(\beta_+ + \sqrt{3}\beta_-,
\beta_+ - \sqrt{3}\beta_-,-2\beta_+\right).$

Assuming that the metric remains diagonal quantum mechanically is equivalent
to solving the classical momentum constraints before quantization \cite{HW95}
({\it cf.\ }\cite[equation E.2.34]{Wald84}); the only remaining gauge freedom
then lies in the time-reparametrizations of the surfaces of homogeneity.  We
shall assume that the metric remains diagonal in the sequel.  However, the
quantum theory thus obtained is not obviously equivalent to a theory in which
the momentum constraint is instead imposed as an operator
condition.\footnote{The issue is that while the momentum constraint implies
that the spatial metric may always be diagonalized at any one moment of time,
the classical equations of motion are required to show that it remains so
thereafter.  These issues are discussed with characteristic lucidity in
section IV of \cite{HW95}.}

Inserting a diagonal homogeneous metric (\ref{eq:twoone}) into the action 
$S = (1/16\pi) \int (R-2\Lambda) + S_M$ and doing the spatial integration over
a standard coordinate volume of $(4\pi)^2$ (the coordinate volume of Bianchi
IX's $SU(2)$ closed spatial manifold; {\it cf.\ }\cite[Box 30.1]{MTW}) yields
the reduced minisuperspace action
\begin{equation}
S= \int dt \left[p_\alpha \dot\alpha + p_+\dot\beta_+ + p_- \dot\beta_- +
p_\phi \dot\phi - \left(\frac{Le^{-3\alpha}}{24\pi}\right) H\right]
\label{eq:twofour}
\end{equation}
where $H$ is the super-Hamiltonian
\begin{equation}
H= -p^2_\alpha + p^2_+ + p^2_- + p^2_\phi + e^{4\alpha} V_\beta
(\beta_+,\beta_-) + e^{6\alpha} V_\phi(\phi) + e^{6\alpha} \Lambda
\label{eq:twofive}
\end{equation}
after rescaling $\phi, V_\phi,$ and $\Lambda$ by positive constants
($\sqrt{4\pi/3}, 192\,\pi^3,$ and $48\,\pi^2$, respectively).
The potential  $V_\beta$ is given in general by
\begin{equation}
V_\beta\left(\beta_+,\beta_-\right) =  
  -24\pi^2\,  e^{2\alpha}\ {}^{(3)}\! R\left(\alpha,\beta_+,\beta_-\right),
\label{eq:twosix}
\end{equation}
where $\,{}^{(3)}\! R$ is the scalar curvature of the homogeneity
hypersurfaces.  Various explicit expressions for $V_\beta$ may be
read off from corresponding expressions in \cite{Mac73,RS75,Wald83,HW95}.  
For the characteristic Bianchi IX (``mixmaster'' \cite{Misner69b}) 
universe that will be the subject of Sections \ref{sec:dffind}\ 
and \ref{sec:semiclassical}\ it is \cite[Box 30.1, for example]{MTW}
\begin{equation}
V_\beta =  12\pi^2\ {\rm tr}\left[e^{4\beta} - 2e^{-2\beta}\right]\ .
\label{eq:twoseven}
\end{equation}
(For ease of comparison with the literature, note that it is traditional to
discuss $V_\beta$ in terms of the anisotropy potential 
$V_a = (1/(6\pi)^2) V_\beta + 1 = 
\frac{1}{3}{\rm tr}\left[1-2e^{-2\beta} + e^{4\beta}\right],$ 
because $V_a$ is positive definite with a global minimum
of 0 at $\beta=0$, and is triangularly symmetric about $\beta=0.$)
The homogeneous isotropic FRW universe may be recovered by setting $\beta = 0$ 
in the equations for Bianchi IX.  For them 
$\,{}^{(3)}\! R = (3/2) e^{-2\alpha}$ and $V_{\beta} = -(6\pi)^2.$  
(Note the usual FRW scale factor $a = 2e^{\alpha}$ after changing from the 
Euler-angle $\sigma^i$ basis to the standard polar coordinates on $S^3.$)

The configuration space for these minisuperspace models is spanned by
the variables $(\alpha, \beta_+, \beta_-, \phi)$.  That is, it is the
superspace of spatial geometries plus the space of scalar field values.
The action (\ref{eq:twofour}) may be expressed in a more compact form by
introducing the notation $q^A, A=0,1,2,3$ for the four variables
$(\alpha, \beta_+, \beta_-, \phi)$, the flat Lorentzian DeWitt metric 
$G_{AB} = {\rm diag} (-1,1,1,1)$, and a rescaled lapse function 
$N(t) = e^{-3\alpha(t)} L(t)/24\pi$. Then
\begin{equation}
S\left[p_A,q^A,N\right] = \int^1_0 dt \left[p_A \dot q^A - NH\right]
\label{eq:twoeight}
\end{equation}
where
\begin{eqnarray}
H&=&G^{AB} p_A p_B + e^{4\alpha} V_\beta (\beta_+,\beta_-) 
    + e^{6\alpha} \left[V_\phi (\phi) + \Lambda\right] \nonumber\\
 &=& G^{AB} p_A p_B + V(\alpha,\beta_+,\beta_-,\phi,\Lambda),
\label{eq:twonine}
\end{eqnarray}
and we have taken advantage of the arbitrariness in $t$ as a coordinate label
to assign the values 0 and 1 to the ends of the range of integration, a choice
we shall make wherever convenient.  In this form the analogy with a
relativistic particle moving in a potential is clear \cite{lesH,HalThor01} and
we shall exploit this in what follows.

The form of the metric (\ref{eq:twoone}) is left unchanged by
reparametrizations of the time $t\to f(t)$.  As already noted,
this invariance is the
remnant of the four-dimensional diffeomorphism invariance of the full theory 
of general relativity once the diagonal form (\ref{eq:twoone}) has been
fixed. Correspondingly the action (\ref{eq:twoeight}) is invariant under
reparametrization transformations of the form
\begin{subequations}
\label{eq:twoten}
\begin{eqnarray}
q^A(t) \to {\tilde q}^A(t) &=& q^A(f(t))\ , \label{eq:twotena}\\
p_A(t) \to {\tilde p}_A (t) &=& p_A (f(t))\ ,  \label{eq:twotenb}\\
N(t) \to {\tilde N}(t) &=& N(f(t)) \dot{f}(t)\ , \label{eq:twotenc}
\end{eqnarray}
\end{subequations}
so long as $f(0)=0$ and $f(1)=1.$

Variation of (\ref{eq:twoeight}) or (\ref{eq:twofour}) with respect to the
multiplier $N$ gives the constraint
\begin{equation}
H\left(p_A, q^A\right) =0
\label{eq:twoeleven}
\end{equation}
between coordinates and their conjugate momenta.  Variation with
respect to the $p_A$ and $q^A$ give Einstein's equations of motion.  The
character of the solutions to these equations --- the possible classical
histories --- depends on the isometry group and the value of the
cosmological constant $\Lambda$. For example, in the case of $\Lambda =
0$ Bianchi I models we expect cosmologies to expand forever from an 
initial singularity.  More precisely, we expect the extrinsic curvature 
$K = (3/L) d\alpha/dt$ (trace of the extrinsic curvature tensor) of the $t=$
constant surfaces of homogeneity to remain positive to the future of an
initially singular surface.
Bianchi I classical solutions with $\Lambda > 0$ also expand forever,
while models with $\Lambda < 0$ always recollapse.
Bianchi IX models have closed spatial sections with three-sphere topology.  
The $\Lambda=0$ Bianchi IX universes are just the
anisotropic generalizations of the closed FRW universe.
More precisely, Lin and Wald \cite{LW90} have shown that when the dominant
energy condition is satisfied and the trace of the spatial projection of the
stress-energy tensor ({\it i.e.\ }the sum of the principal pressures) is
positive -- thus excluding a cosmological constant -- there are no classical
solutions which expand forever from an initial singularity in the sense that
the extrinsic curvature of constant $t$ surfaces remains positive.  

The stress-energy tensor arising from (\ref{eq:twothree}) satisfies the
dominant energy condition so long as $V_{\phi}\geq 0$, though the pressures
may be negative if the potential energy in the scalar field exceeds the
kinetic energy.  (Indeed, it is precisely this feature which allows a scalar
field to mimic a cosmological constant in inflationary models.)  Thus, scalar 
fields only satisfy the conditions of the Lin-Wald recollapse theorem at late 
times for certain 
choices of scalar potential -- a free, massless, minimally coupled scalar
field, for example.
The general conditions on $V_{\phi}$ for which a recollapse theorem holds are 
as far as we are aware not currently known.

With a cosmological constant the conditions of the Lin-Wald theorem are not
satisified.  The example of de Sitter space is enough to show that there will
be Bianchi IX solutions with vanishing scalar field which evolve
non-singularly from an initial contracting phase $(K<0)$ in the infinite past
to an expanding phase $(K>0)$ in the far future.  In between the volume of the
universe reaches a non-zero minimum value.  The inclusion of a small amount of
homogeneous scalar field would not be expected to disturb this behavior.
However, the homogeneous, isotropic Bianchi IX (Friedmann) models show that a
positive cosmological constant does not remove the singularity in every case.
There are also models in which the stress-energy of the scalar field dominates
that of the cosmological constant throughout the model's history.  Such models
display the qualitative features of the $\Lambda=0$ case --- an initial
singularity leading to a finite expansion followed by recontraction, for 
models satisfying the conditions of the Lin-Wald theorem.  See
\cite{WE97} for a wide-ranging survey of the dynamics of the various
cosmological models.

\section{Generalized Quantum Mechanics of $\Lambda=0$ Bianchi IX  
Universes}
\label{sec:IXLambdaeq0}

\subsection{Generalized Quantum Theory}
\label{sec:gqt}

In this section we describe a generalized quantum theory for Bianchi IX
minisuperspace cosmological models with $\Lambda=0$.  While the construction
is in principle valid even for models which may expand forever, certain
technical aspects of the analysis are simplified by restricting attention to
quantum histories which always recontract, and the most general case will not
be considered here.

We work within the general principles of the quantum mechanics of a closed
system \cite{Gri84,Omnsum,GH90a}.  The closed system is most generally and
accurately the universe as a whole.  The most general predictions of quantum
mechanics are the probabilities of individual members of sets of alternative
coarse-grained histories of the closed system.  Probabilities are not
predicted for every set of alternative histories, but only those for which the
quantum mechanical interference between the individual histories in the set
is negligible as a consequence of the system's boundary conditions and
dynamics.  Such sets of histories are said to decohere, or be consistent.

Generalized quantum theory \cite{lesH,Ish94,Har91a} is a comprehensive
framework for implementing the principles of the quantum mechanics of closed
systems.  As noted in the introduction, the following elements specify a
generalized quantum mechanics:
(1) The sets of fine-grained histories which are the most refined
description of the system possible. (2) The allowed coarse-grainings
which generally are partitions of a set of fine-grained histories into
an exhaustive set of mutually exclusive classes $\{c_h\}, h =
1,2,3 \cdots$ called coarse-grained histories. (3) A decoherence
functional, $D(h,h^\prime)$ that measures the interference
between pairs of histories in a coarse-grained set.  The decoherence
functional is a complex-valued functional on pairs of classes that 
satisfies certain general requirements:  It is (i) Hermitian, (ii)
normalized, (iii) positive, and (iv) consistent with the principle of
superposition in senses made precise in \cite{lesH,Ish94,Har91a}. The
decoherence functional incorporates a specification of the boundary
conditions for the closed system --- typically ``initial'' and ``final''
conditions.  It is a natural generalization to closed systems of the
idea of quantum state, as the term is used in quantum logic and in
algebraic quantum mechanics, to measure the quantum interference
between histories in addition to their probabilities \cite{IL94,dac97}.

With these three elements specified, the process of prediction proceeds as
follows: A set of alternative coarse-grained histories (approximately, medium)
decoheres when $D(h,h^\prime)$ is negligible for all $h\not= h^\prime$.  The
probabilities $p(h)$ of the individual histories in a decoherent set are the
diagonal elements of $D$.  The rules for decoherence and probabilities are
thus summarized by the fundamental formula
\begin{equation}
   D\left(h^\prime,h\right) \approx p(h)\, \delta_{h^\prime h}  
\label{eq:threeone}
\end{equation}
obeyed by histories in decohering sets.

The four-dimensional diffeomorphism invariance characteristic of a geometric
theory of gravity is most easily accomodated by employing a sum-over-histories
formulation of quantum mechanics.  A sum-over-histories quantum mechanics
posits a unique set of fine-grained histories which in the case of gravity are
four-dimensional spacetime metrics and matter field configurations.  A
sum-over-histories generalized quantum mechanics for cosmology was described
in \cite{lesH}.  Generalized quantum mechanics for models with a single
reparametrization invariance was described as well.%
\footnote{The generalized quantum mechanics for cosmology in \cite{lesH} is
but one of many applications of sum-over-histories principles and techniques
to quantum gravity.  These are too numerous and familiar to be cited here, but
two particular applications to quantum cosmological dynamics should be
mentioned whose spirit is consonant with that in \cite{lesH}.  C.~Teitelboim
\cite{Teiall} developed a sum-over-histories ``$S$-matrix'' theory for
calculating transition amplitudes between alternatives defined near
cosmological singularities.  We exploit the formal techniques developed by
Teitelboim for constructing functional integrals over spacetime geometries and
in particular his choice for the measure in such integrals.  The way
probabilities are related to amplitudes differs and we aim, in a generalized
quantum mechanical setting, at probabilities for much more general and
accessible alternatives than those that can be defined near singularities. 
R.~Sorkin's \cite{Sor94} treatment of general alternatives is concordant with
that of the present work.
} %
We now apply those discussions to homogeneous, minisuperspace cosmological
models which, as formulated here, possess a single time reparametrization
invariance.  To do that we specify explicitly the three elements of their
generalized quantum mechanics.

\subsubsection{Fine-grained Histories}  
\label{sec:fighs}  
We take for the set of fine-grained histories the paths 
$z^M(t) = (q^A(t), N(t))$, $M = 0,\ldots,4$  
in the extended configuration space ${\cal C}_{\rm ext}$ of $q^A=(\alpha,
\beta_+, \beta_-, \phi)$ and multiplier $z^4 = N$.  We are thus considering a
sum-over-histories quantum mechanics in which there is a unique fine-grained
set of histories.  We put no restriction of single-valuedness on the paths. 
Thus, for example, the total spatial volume $(4\pi)^2 {\rm exp} (3\alpha(t))$
may increase and decrease over the course of the history and, indeed, go
through such cycles an arbitrarily large number of times.  In particular, the
classical histories of a Bianchi IX universe which expand and recontract are
included among the possible quantum mechanical histories.

We put no restriction on the differentiability of the paths, but do require
them to be continuous.

The fine-grained histories have ends at which the cosmological boundary
conditions analogous to initial and final conditions are imposed, and these
ends must be prescribed to complete the specification of the set of
fine-grained histories.  A natural principle restricting this choice is that
{\it the set of fine-grained histories should include all the classical
histories}.  Otherwise there is no hope of recovering Einstein's classical
theory as a suitable limit of quantum theory.  (This will be the subject of
Section \ref{sec:semiclassical}.)  The work of Lin and Wald \cite{LW90}
discussed at the end of section \ref{sec:models} shows
that all classical Bianchi IX cosmologies with scalar field and $\Lambda=0$
expand from an initial singularity of vanishing three-volume and eventually
recontract to a final singularity of vanishing three-volume, assuming that the
potential energy in the scalar field does not dominate the kinetic at late
times.  We shall confine our attention to potentials for which this is so.
Classical histories may therefore be thought of as beginning and ending on a
surface $\sigma_0$ of large constant negative $\alpha (=\alpha_0)$ in the
configuration space of paths.  The class of all paths that begin and end on
such a surface is therefore the natural, minimal set of fine-grained histories
for a generalized quantum theory of Bianchi IX minisuperspace cosmological
models with scalar field and $\Lambda=0$.  
This was the choice advocated by Teitelboim in his theory of quantum
cosmological ``scattering'' between initial and final singularities
\cite{Teiall}.  It is the choice we shall adopt here.  (Restricting the paths
in this way corresponds to the imposition of a boundary condition that wave
functions vanish as $\alpha \rightarrow \infty$.)  Finally, we add the
restriction that all paths possess $\dot{\alpha}(0) > 0$ and $\dot{\alpha}(1) <
0$ on $\sigma_0$, {\it i.e.\ }``expanding'' initial and ``contracting'' final
conditions.

\subsubsection{Coarse-grained Histories}  
\label{sec:coghs}  

Coarse-grained histories correspond to the physical questions that
may be asked of a system.  We therefore allow as coarse-grained
sets of alternative  histories any partition of the
fine-grained histories into {\it reparametrization invariant} classes
$\{c_h\},\ h=1,2,\cdots$, because reparametrization invariance is what 
remains in minisuperspace of the diffeomorphism invariance of
general relativity.  The classes generally may be thought of as
partitions by values of reparametrization invariant functionals 
$F[q^A,N]$ of the paths in ${\cal C}_{\rm ext}$. Explicitly, for an 
exclusive set
of ranges $\{\Delta_h\}, h = 1,2,\cdots$ of the real line, and
a single functional $F$ we define
\begin{equation}
c_h = \left\{(q^A(t), N(t))\ | \ F[q^A,N] \in \Delta_h\right\}\ .
\label{eq:threetwo}
\end{equation}
Any partition may be thought of as of this form because we may always
consider the functional which has the value $h$ for paths in
$c_h$ and a set of ranges that bracket the integers.

Simple examples of interesting partitions into diffeomorphism invariant
classes -- here reduced to reparameterization invariant
classes -- are readily given:

\begin{itemize}

\item One could partition the histories by ranges of values of the volume of
the largest volume three surface of homogeneity.  The resulting
probabilities are for the values of the volume of maximum expansion of the
universe reached in the course of its expansion and contraction.

\item One could partition the histories of these homogeneous space times
into the classes which have a surface of homogeneity with a volume less
than $\epsilon$ and the class of those with no such surface. The value of the
probability that there is a surface with volume less than $\epsilon$
considered as $\epsilon\to 0$ would be one way of assigning a probability to
the universe becoming singular.%
\footnote{There could be many other meanings assigned to a singular quantum
evolution, for example, a finite probability for a curvature invariant to have
a divergent value ({\it cf.\ }\cite{HM95}).  The various possible definitions
are not necessarily easily related or of equal significance because the paths
are non-differentiable.  The particular criterion for singular evolution ---
finite probability for a surface of homogeneity of vanishing volume --- that is
under discussion for these minisuperspace models would not be a valid criterion
for a full theory of quantum cosmology for two reasons: First a general
geometry may have no surfaces of homogeneity.  Second, spacelike surfaces of
nearly vanishing volume may be constructed even in non-singular spacetimes from
segments that are nearly null.  The criterion is meaningful in these
minisuperspace models only because {\it all} fine-grained histories can be
foliated by surfaces of homogeneity --- by assumption.  The important point,
however, is that any meaningful criterion for singular quantum evolution that
is expressible in spacetime form corresponds to a partition of the fine-grained
histories into the class which is singular by that criterion and the class
which is not singular.  If the alternatives of \{(singular), (non-singular)\}
{\it decohere}, then a generalized quantum theory will assign a probability to
a quantum universe being singular.  Unlike the classical theory, however, {\em
a finite probability for singular geometry need not mean a breakdown of
predictability.}
} %

\item One could partition the histories into the class which remain 
close to a solution of the classical Einstein equation by some standard 
({\it cf.\ }(\ref{eq:distanceF})) and the
class which exhibit a significant excursion away from classical
behavior.  The probability of the first class is the probability that
the universe behaves classically according to the given standard. We
shall employ such coarse grainings in Section \ref{sec:semiclassical}.

\item One could partition the fine-grained histories by ranges of values of
the volume and anisotropy $\beta_{\pm}$ they assume a given ``proper
time''\footnote{We follow terminology of Teitelboim \cite{Teiall} and Henneaux
and Teitelboim \cite{HT83} in calling the reparametrization invariant quantity
$\int Ldt$ the ``proper time''.  K.~Kucha\v r \cite{KKpriv} has stressed that
this may be confusing and suggests the term ``separation''.}
$\int Ldt$ after the initial condition.  The resulting probabilities would be
the probabilities for the volumes and anisotropies the universe could have at
a given proper time from the initial surface.
These are not unlike 
the probabilities that would be of interest in comparing the predictions of a
realistic quantum cosmology with observation.

\end{itemize}

All of the coarse grainings mentioned above are into manifestly
reparametrization invariant classes of fine-grained histories.  Most are
partitions that are not defined by alternatives that are in any sense
alternatives ``at one moment of time'', and hence are not defined simply by
observables on superspace.  Rather they are spacetime alternatives
referring to properties of histories extended over time.  For example, the
probabilities for classical behavior refer to whether a suitably
coarse-grained history approximately obeys the Einstein equation over a course
of time.

Coarse-grained histories correspond to the physical questions that may be
asked of a system in the following sense.  Ask for the probability that the
universe has {\it any} reparametrization invariant property expressible in
spacetime terms.\footnote{In path integral formulations of this kind, physical
quantities that involve {\it derivatives} such as the extrinsic curvature $K$
must be expressed in terms of differences of metric variables at different
times.}
To answer this question one considers the partition of the fine-grained
histories into the class which have the property and the class which do not. 
If this set of coarse-grained histories decoheres, then the quantum mechanics
predicts the probability that the universe has the property in question.  If
one cannot tell whether a given fine-grained history has the property or not
then either it does not make sense or it is not expressible 
in terms of metric and/or matter field configurations alone.

While it is easy to exhibit physically interesting sets of alternative
reparametrization invariant coarse-grained histories for these homogeneous
minisuperspace cosmological models it is much harder to find sets of such
histories that decohere.  That is because of the small number of degrees of
freedom of the model.  Coarse-graining is essential for decoherence. 
Realistic mechanisms of decoherence that are effective in a variety of initial
conditions qualitatively involve the dissipation of phases from variables
followed by the coarse-graining into variables that are ignored.\footnote{See,
e.g.~\cite{DH92,GMH93} for more quantitative discussions from a histories
point of view, as well as \cite{Giulini}.} However, the present minisuperspace
models do not present many degrees of freedom to be ignored!

\subsubsection{The Decoherence Functional}  
\label{sec:df}
The decoherence functional is constructed in two steps following the analogy
of the relativistic particle discussed in \cite{lesH}; the final result is
expressed in (\ref{eq:threenine}).  It will turn out to be a natural
generalization to a reparametrization invariant theory of the canonical
decoherence functional of ordinary quantum mechanics.

First we define matrix elements of class operators corresponding to individual
coarse-grained histories $h$ by a sum-over-histories in the class. 
Schematically, we define
\begin{equation}
\left\langle q'' \| C_h \| q' \right\rangle = 
\sum\limits_{{\rm paths}\ \in \bigl[q' h q^{''}\bigr]}
\hspace{-6pt}  \exp\bigl(i\, S[{\rm path}]\bigr)
\label{eq:threethree}
\end{equation}
where the sum is over paths in ${\cal C}_{\rm ext}$ that lie in the class
$c_h$ which begin at $q'$ and end at $q''$.  $S$ is the (Lagrangian) action
for paths --- a functional of $q(t)$ and $N(t)$.
In order to give a definite meaning to the functional integral in 
(\ref{eq:threethree}), and, in particular, to fix the ``measure'' on the space
of paths, it is convenient to consider the corresponding integral over
paths in phase-space,
\begin{equation}
\left\langle q'' \| C_h \| q' \right\rangle =
 \int_{c_h} \delta N_{\geq 0}\, \delta p\, 
      \delta q^{q^{\prime\prime}}_{q^{\prime}}\, 
   \Delta_\Gauge [q,N]  \delta \bigl[\Gauge [q,N]\bigr] 
    \exp \biggl(i\, S\bigl[p_A,q^A,N \bigr]\biggr)
\label{eq:threefour}
\end{equation}
The ingredients in this expression are as follows: $S$ is the action
(\ref{eq:twoeight}).  $\Gauge$ is a function such that the condition
$\Gauge=0$ fixes a unique representative from each reparametrization invariant
equivalence class of fine-grained histories; $\Delta_\Gauge$ is the associated
Faddeev-Popov determinant.  The integral is over all paths in ${\cal C}_{\rm
ext}$ that lie in the class $c_h$ and pass between the configuration space
points $(q^A)'$ and $(q^A)''$.  All possible momentum paths are integrated
over.  We take the multiplier functional integral to be over positive values
$N\geq 0$.  Other choices, for example, both positive and negative values,
would lead to different generalized quantum theories.  (Some of the issues
that arise in choosing the allowed range for the lapse are discussed in
\cite{Teit83a,HTV92,Hal88,HH91}
and \cite[section VII]{lesH}.)  The measure for the $q$ and $p$ integrations
is the usual Liouville
``$dqdp/2\pi$'' measure on phase-space paths.  The integration over momenta
may be seen as a device to induce the measure on paths in ${\cal C}_{\rm ext}$
from the Liouville measure since the range of integration over the momenta
are constrained neither by the class $c_h$, the endpoints $(q', q'')$, or the
gauge fixing delta-function.  The skeletonized path integrals we envision are
quite standard \cite{Teiall,Hal88}, and have already been briefly described in
\cite{lesH}.

The decoherence functional is constructed from the class operators
(\ref{eq:threefour}) which characterize the physical histories in question,
but also incorporates a theory of cosmological boundary conditions imposed at
the ends of the histories that are the analogs of initial and final conditions
in a quantum system with a fixed notion of time.

An initial or final condition is represented by a set of configuration-space
wave functions together with a probability for each wave function.  That is
the same kind of information needed to specify an initial or final density
matrix in ordinary quantum mechanics.  The wave functions are required to
satisfy an operator implementation of the constraints.  Thus, for homogeneous
cosmological models an initial condition is specified by a set $\{\Psi_i
(q^A), p^\prime_i\}$, where each $\Psi_i(q^A)$ satisfies an operator form of
(\ref{eq:twoeleven}).  We take the obvious operator ordering%
\footnote{Most generally, the operator ordering in the Wheeler-DeWitt equation
obeyed by the boundary wave functions should agree with the operator ordering
in the Wheeler-DeWitt equation obeyed by the full propagator, (\ref{eq:aone}).
This ordering in turn is determined by the precise form of the measure in the
functional integral (\ref{eq:threefour}) defining the class operators. 
In the present instance, however, we have effectively circumvented this issue
at the classical kinematical level by restricting attention to an action of
the form (\ref{eq:twonine}) expressed in coordinates in which the kinetic
term is independent of the minisuperspace coordinates; the operator ordering
appropriate to this choice is just that of (\ref{eq:threefive}).
} 
and write
\begin{equation}
\label{eq:threefive}
H\, \Psi_i(q) = 
   \left[-G^{AB}\, \frac{\partial^2}{\partial q^A\partial q^B} +
   V (q)\right]\Psi_i(q) = 0\ .
\end{equation}
Similarly a final condition is specified by a set 
$\{\Phi_j(q^A), p^{''}_j\}$, where the
$\Phi_j(q^A)$ also solve the Wheeler-DeWitt equation (\ref{eq:threefive}).  
In order to guarantee the independence of the decoherence functional from the
choice of initial surface $\sigma_0$ -- see below (\ref{eq:threeeight}) -- and
for additional reasons that will be explained in the sequel, 
we require the initial solutions $\Psi(q)$ of the Wheeler-DeWitt equation to
decay in ``spatial'' directions in minisuperspace (that is, on surfaces of
constant $\alpha$) fast enough in $\beta_{\pm}$ and $\phi$ that $V\Psi \approx
0$ as $\alpha \rightarrow -\infty$.  (More precisely, we require that
$e^{-4\alpha}V\Psi$ is bounded; see section \ref{sec:findifference}.)  The
space of functions on minisuperspace with these properties will be denoted
$\Fall(\sigma_0).$ (\ref{eq:threefive}) is the Wheeler-DeWitt equation for the
``initial'' and ``final'' conditions in this minisuperspace quantum cosmology.
 
At this stage in the construction of the decoherence functional, the
specification of the minisuperspace boundary conditions is otherwise
essentially arbitrary.  Ultimately, these boundary conditions must be 
supplied by a detailed physical {\it theory} of the cosmological
boundary conditions.  \cite{Hal91,Wiltshire97} are
reviews of this aspect of quantum cosmology.

Initial and final conditions are adjoined to the class operator matrix
elements by means of a linear product $\circ$ on the 
space of solutions to (\ref{eq:threefive}). Following \cite{lesH} we 
shall take for $\circ$ the Wheeler-DeWitt (Klein-Gordon) product.  
Specifically, given a surface $\sigma$ in the
configuration space that is spacelike in the metric $G^{AB}$ we define
\begin{equation}
\Phi \circ \Psi = i\int\nolimits_\sigma d \Sigma^A\ \Phi^*(q)
\buildrel\leftrightarrow\over{\nabla}_{\kern-0.2em A} \Psi(q).
\label{eq:threesix}
\end{equation}
The inner product, so defined, is independent of the surface used to
define it so long as $\Phi$ and $\Psi$ both satisfy
(\ref{eq:threefive}).
In the coordinates $q^A = (\alpha, \beta_+, \beta_-, \phi)$ in which $G^{AB}
= {\rm diag} (-1,1,1,1)$ a convenient choice of $\sigma$ is typically a
surface of constant $\alpha$, whence
\begin{equation}
\Phi\circ\Psi = i \int\nolimits_\alpha d^3 q\ \Phi^* (\alpha, \vec{q})
\ \frac{\buildrel\leftrightarrow \over\partial}{\partial\alpha} 
\Psi (\alpha, \vec{q})\ .
\label{eq:threeseven}
\end{equation}

Having introduced the product $\circ$ we now define
\begin{equation}
\left\langle\Phi_i | C_h | \Psi_j \right\rangle =
\Phi_i \left(q'' \right) \circ \left\langle q'' \| C_h \|
q' \right\rangle \circ \Psi_j \left(q'\right)\ .
\label{eq:threeeight}
\end{equation}
This definition appears at first sight to depend on the choice of surface on
which $\circ$ is defined, but in fact it does not, so long as the coarse
graining $c_{h}$ does not restrict the paths on those surfaces.  This is
because the class operators (\ref{eq:threefour}) for such coarse grainings
generally solve (\ref{eq:threefive}) as well as $\Psi$ and $\Phi$, a fact we
demonstrate in the appendix.  First, however, we complete the definition of
the generalized quantum theory of homogeneous minisuperspace cosmologies.

The decoherence functional $D(h^\prime,h)$ is defined through
\begin{equation}
D(h^\prime, h) = {\cal N} \sum_{i,j} p^{''}_i
\left\langle \Phi_i | C_{h^\prime} | \Psi_j \right\rangle p^\prime_j 
\left\langle\Phi_i | C_h | \Psi_j\right\rangle^*
\label{eq:threenine}
\end{equation}
where ${\cal N}$ is a normalizing factor determined so that
$\Sigma_{h^\prime h} D(h^\prime, h)=1$. 
Specifically, if $u$ is the class of {\it all} fine-grained histories
\begin{equation}
     {\cal N}^{-1} = \sum\nolimits_{ij} p^{''}_i | \left\langle\Phi_i | C_u |
\Psi_j\right\rangle |^2 p^\prime_j\ .
\label{eq:threeten}
\end{equation}

The decoherence functional defined by (\ref{eq:threenine}) is the direct
analogue of the ``canonical'' decoherence functional of ordinary Hamiltonian
quantum mechanics with initial and final boundary conditions
\cite{lesH,dac97}, written in functional integral form, with accomodations
appropriate to the reparametrization invariance of the present theory.  It
satisfies the general conditions required of a generalized quantum theory.  It
is (i) Hermitian, $D(h^\prime,h) = D(h, h^\prime)^*$, (ii) normalized,
$\Sigma_{h h^\prime} D(h^\prime,h)=1$, (iii) positive on the diagonal
elements, $D(h,h)\geq0$, and, (iv) consistent with the principle of
superposition in the sense that if $\{\bar c_{\bar h}\}$ is a partition of the
classes $\{c_h\}$ into coarser classes, then
\begin{equation}
    D(\bar h^\prime, {\bar h}) = \sum\limits_{h^\prime\in
\bar h^\prime} \sum\limits_{ h\in\bar h} D(h^\prime, h)
\label{eq:threeeleven}
\end{equation}
These four conditions are enough to ensure that for sets of histories that
decohere according to (\ref{eq:threeone}), the numbers $p(h)$ defined by
(\ref{eq:threeone}) are probabilities satisfying the most general form of the
probability sum rules.  By using in (\ref{eq:threeone}) the specific form
(\ref{eq:threenine}) we can assess the probabilities of alternative,
coarse-grained, decohering histories of the model homogeneous cosmologies
under discussion.

\subsection{Evaluation of the Class Operators in the
Proper Time Gauge} 
\label{sec:Ndot}

We apply (\ref{eq:threeone}) to predictions concerning the semiclassical
behaviour of homogeneous minisuperspace cosmologies in the next section.  To
end this section, we discuss the evaluation of the class operators
(\ref{eq:threefour}) in a particularly convenient gauge --
(\ref{eq:threethirteen}) -- called the ``proper time'' gauge.  For suitable
coarse grainings, we also briefly argue that these matrix elements satisfy the
constraint (\ref{eq:threefive}).
Because the action (\ref{eq:twoeight}) is essentially that of a
relativistic particle in a potential, the treatment closely parallels
that of the free relativistic particle that has been given previously
in \cite[section 7]{lesH}.

The first step in evaluating (\ref{eq:threefour}) is to choose a ``gauge" that
fixes the reparametrization symmetry (\ref{eq:twotena}-\ref{eq:twotenc}), the
infinitesimal form of which is invariance under the changes (set $f(t) = t +
\epsilon/N$)
\begin{subequations}
\label{eq:threetwelve}
\begin{eqnarray}
\delta q^A &=& \epsilon(t)\{ q^A,H \}  \label{eq:threetwelvea}\\
\delta p_A &=& \epsilon(t)\{ p_A,H \}  \label{eq:threetwelveb}\\
\delta N\  &=& \dot{\epsilon}(t),      \label{eq:threetwelvec}  
\end{eqnarray}
\end{subequations}
where $\{\ ,\ \}$ is the Poisson bracket, and $\epsilon(0) =
\epsilon(1) = 0$.  A convenient ``gauge" fixing function is
\cite{Teit83a,HTV92} (see also \cite[section 7]{lesH})
\begin{equation}
\label{eq:threethirteen}
  \Gauge = \dot{N}.
\end{equation}
The Fadeev-Popov determinant 
$\Delta_{\Gauge} = {\rm det}\, \delta\Gauge/\delta\epsilon \sim{\rm det}\, d^2/dt^2$ 
is in this case independent of the integration variables.  The delta
functional in (\ref{eq:threefour}) then permits only the $N= {\rm constant}$
paths to contribute to the integral over $N(t)$, leaving
\begin{equation}
\label{eq:threefourteen}
\Melt{q''}{C_h}{q'} =
   \int_{c_h}dN_{\geq 0}\, \delta p\, \delta q_{q'}^{q''} 
   \exp\left( i \int_0^1 dt\, [ p_A\dot{q}^A - NH ] \right)
\end{equation}
after dropping the constant factors that cancel in the decoherence
functional (\ref{eq:threenine}).  Changing variables to
\begin{equation}
\label{eq:threefifteen}
ds = N\, dt,
\end{equation}
the Gaussian functional integrals over $p$ may be performed, leaving simply
\begin{equation}
\label{eq:threesixteen}
\Melt{q''}{C_h}{q'} = \int_0^{\infty} dN\, \Melt{q''\, N}{C_h}{q'\, 0}
\end{equation}
so long as we assume the coarse graining does not restrict the value of 
$N$; otherwise the range of the $N$ integral must be restricted 
appropriately as well.  Here we have defined
\begin{subequations}
\label{eq:threeseventeen}
\begin{eqnarray}
\Melt{q''\, N}{C_h}{q'\, 0} &=&
   \int_{c_h} \delta q _{q'}^{q''} \exp\left( i \int_0^1 dt\, 
   \left[ \frac{1}{4N}G_{AB}\dot{q}^A\dot{q}^B - NV \right] \right)  
        \label{eq:threeseventeena}\\ &=& 
   \int_{c_h} \delta q _{q'}^{q''} \exp\left( i \int_0^N ds\, \left[ 
   \frac{1}{4}G_{AB}\frac{dq^A}{ds}\frac{dq^B}{ds} - V \right] \right),
        \label{eq:threeseventeenb}
\end{eqnarray}
\end{subequations}
where 
the path integral measure $\delta q$ has been renormalized in the usual manner
induced by the momentum integrations.  (See the appendix for details.)  The
notation on the left hand side of (\ref{eq:threeseventeena}) is inspired by
the observation that the path integral in (\ref{eq:threeseventeenb}) shares
the form of that for the propagator over a time $N$ of a relativistic particle
in a potential $V$. 
(\ref{eq:threesixteen}) may be thought of as a ``restricted propagator'' 
for the class of paths $c_h,$ as should be evident from the restricted
functional integral (\ref{eq:threefour}).

In general, the complexity of the minisuperspace potential $V$ precludes much
further explicit progress in the non-perturbative evaluation of
$\Melt{q''}{C_h}{q'}$.  It is still possible, however, to show that the class
operators satisfy the constraint (\ref{eq:threefive}) for coarse grainings
which do not restrict the values of the endpoints $q'$ and $q''$ or the value
of $N$.%
\footnote{Another approach is to {\it require} that the class operators
satisfy the constraints and to modify their definition appropriately
\cite{HalThor01,HM97,HalThor02,Hal04}.  We do not expect such modifications to
fundamentally affect any of the specific results presented here, and thus
defer their consideration.
} %
This is done in the appendix.  (\ref{eq:threeeight}) is thus as
claimed independent of the surfaces on which we choose to impose the boundary
conditions $\Psi$ and $\Phi$.

\section{Decoherence Functional For Recollapsing Bianchi Cosmologies}
\label{sec:dffind}

The general form of the decoherence functional for type A homogeneous 
cosmologies is given in (\ref{eq:threenine}).  In this section we employ some 
specific choices of the initial and final conditions approriate to $\Lambda = 
0$ closed (Bianchi IX) cosmologies to cast the decoherence functional into a 
simpler and more practically useful form.

\subsection{Initial and Final Conditions}
\label{sec:bc}

As already noted, it is to be expected that the initial quantum conditions of
the universe are fixed by some 
{\it theory} of cosmological boundary conditions.  Knowledge of this theory is
not, however, required in the construction of the decoherence functional for
cosmology.  We shall, therefore, illustrate the process of prediction based on
the decoherence functional (\ref{eq:threenine}) with some simple choices of
initial state.  The practical significance of such predictions depends
entirely on whether the chosen initial states are representative of the
boundary conditions on the actual physical universe.

Most of the extant theories of the initial state of the universe
\cite{Hal91,Wiltshire97}\ produce a boundary state consisting in a single
initial wave function.  We will therefore in our examples concentrate entirely
on the case of a pure initial state $\Psi.$

\begin{figure}[!htbp]   
\includegraphics[scale=0.75]{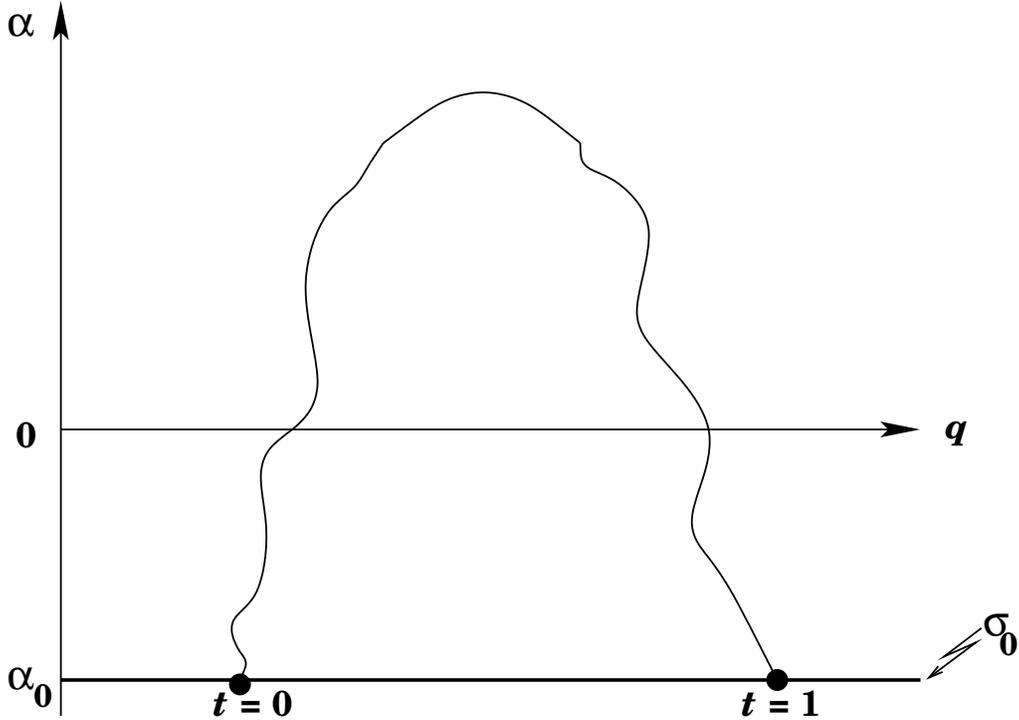}
\caption{\label{fig:fighs}  
The minisuperspace of homogeneous cosmological models.  The timelike
coordinate $q^0 = \alpha$ and one spacelike coordinate ({\it e.g.\ }$q^1 =
\beta_+$) are plotted.  Fine grained histories are paths in this configuration
space which begin and end on a surface $\sigma_0$ that is spacelike in the
Wheeler-DeWitt minisuperspace metric $G_{AB}$ and located at large negative
$\alpha$ ($=\alpha_0$, say), corresponding to a small spatial volume.  
The restriction to paths which begin and end on $\sigma_0$ corresponds to the 
imposition of a boundary condition that solutions to the Wheeler-DeWitt equation 
vanish as $\alpha \rightarrow \infty$.
The paths all possess $\dot{\alpha}(0) > 0$ and $\dot{\alpha}(1) < 0$ on
$\sigma_0$, {\it i.e.\ }``expanding'' initial and ``contracting'' final
conditions.  The universe expands from a small to a maximum volume and then
recontracts.
}%
\end{figure}

Recall from Section \ref{sec:fighs} that the Bianchi IX cosmological histories
have ``ends" at which we impose boundary conditions $\{ \Psi_i, p_i' \}$ and
$\{ \Phi_i, p_i'' \}$.  In order to correspond to a conventional notion of
cosmological boundary conditions for closed, $\Lambda = 0$ universes, we
impose the boundary conditions on a suitable surface $\sigma_{0}$ of large,
negative $\alpha$ ($=\alpha_0$), {\it i.e.\ }when the universe is very small. 
Because the wave functions $\Psi_i$ and $\Phi_i$ satisfy the constraint
(\ref{eq:threefive}), it does not matter on which surface they are imposed so
long as the coarse grainings under consideration do not involve regions of
minisuperspace intersecting those surfaces, as noted above.
In this sense, then, the cosmological histories 
``begin" and ``end" at small spatial volume, just as the classical histories do.  
(See Figure \ref{fig:fighs}.)

Maintaining a close correspondence with ordinary quantum mechanics, we shall
choose final boundary conditions which are ``indifferent'' to the paths, in a
sense to be made precise in Section \ref{sec:findifference}.  There will then
effectively be no final conditions at all on the cosmological histories.

As for the choice of the initial boundary conditions, we will examine two
instructive examples in the sequel.  In order to illustrate how the present
sum-over-histories construction can predict semiclassical behaviour in a
suitable limit as naturally as sum-over-histories formulations always do, as a
first example we consider a single initial localized Wheeler-DeWitt wave
packet $\Psi^{\mathit{WP}}$ which is designed to prefer a particular
classical path over all others in the semiclassical limit.  In a partition of
the minisuperspace histories by classical paths, a steepest descents
evaluation of the path integral for the class operators then reveals that the
primary contribution to the decoherence functional comes from the class
operator corresponding to the coarse-grained class of paths containing the
classical path preferred by the initial condition $\Psi^{\mathit{WP}}.$

Next, we shall turn our attention to the semiclassical predictions of the
decoherence functional with an initial state of the more traditional WKB form. 
As is to be expected, under semiclassical coarse grainings such a choice of
initial state leads to a distribution of classical trajectories with initial
value weighted according to the WKB prefactor, and initial momentum given by
the gradient of the phase.  

Let us now proceed with the details.

\subsection{Branch Wave Functions}
\label{sec:branch}

First, it is useful to define the ``branch wave functions'' 
\begin{eqnarray}
   \Psi_h(q) &=& \langle q \| C_h | \Psi \rangle \nonumber\\
             &\equiv& \langle q \| C_h \| q' \rangle \circ \Psi(q').
\label{eq:branchwavefns}
\end{eqnarray}
These may be regarded as the wave functions corresponding to the
initial state, propagated by the restricted propagator (class operator)
corresponding to the physical history in question.\footnote{It may be 
helpful to be reminded that the ``matrix'' appearing in 
(\ref{eq:branchwavefns}) should not be complex conjugated when 
computing the product.}

The space $\{\Psi_{h}(q)\}$ of branch wave
functions will be denoted $\Branch$.  It depends in an essential way both on
the allowed space of initial wave functions $\Psi(q),$ and on the allowed
coarse grainings.

For arbitrary coarse grainings, it is not immediately evident that
$\Psi_{h}(q)$ for $q \in \sigma_{0}$ must be in the space $\Fall(\sigma_{0})$
of functions which fall off 
rapidly at infinity on $\sigma_{0}$, even when the initial wave functions
$\Psi(q)$ are required to be in $\Fall(\sigma_{0}),$ as we have done. 
However, we will take the arguments of Higuchi and Wald \cite{HW95} as
strongly suggestive that initial states in $\Fall(\sigma_{0}),$ propagated by
the full propagator $\Melt{q''}{C_{u}}{q'}$ off of $\sigma_0$ and then back to
$\alpha\rightarrow -\infty,$ indeed remain in $\Fall(\sigma_{0})$ when there
is a scalar field present.%
\footnote{The essence of their argument is that for scalar potentials which do
not grow exponentially at infinity, the scalar potential term in
(\ref{eq:threefive}) becomes negligible as $\alpha\rightarrow -\infty$ and
hence the scalar momentum acts as a conserved mass term in the equation of
motion.  This means that wave packets ``lose energy'' upon each bounce off of
mixmaster's potential walls, eventually slowing to the point where they move
more slowly than the walls recede.  Wave packets therefore should become
asymptotically ``free'' as $\alpha\rightarrow -\infty$ inside the potential
walls, escaping to infinity more slowly than the receding walls.
} %
For the present, we will assume when necessary that attention is 
restricted to coarse grainings for which all 
$\Psi_{h}(q) \in \Fall(\sigma_{0})$ when $q \in \sigma_0$ 
{\it i.e.\ }for which $\Branch|_{\sigma_0} \subset \Fall(\sigma_{0}).$

\subsection{Final Indifference}
\label{sec:findifference}

Boundary conditions of ``final indifference'' should amount, in
essence, to no final boundary condition at all, in the sense that
the endpoints of all paths are weighted equally.  The final boundary
condition then effectively disappears from the decoherence functional,
just as it does in ordinary quantum mechanics.

Let us be slightly more precise about how to do this.

What we seek are a complete set of solutions to the Wheeler-DeWitt
equation (\ref{eq:threefive}) which may serve to define a positive
``resolution of the identity'' in the space of branch wave functions
$\Branch.$  By this we mean the following.  First, note that if we define
\begin{equation}
     \scri(q''_2,q''_1) 
         = \sum_{i}\ p_{i}''\, \Phi_{i}(q''_2)\Phi_{i}^{*}(q''_1)
\label{eq:scri}
\end{equation}
for $q''_1, q''_2 \in \sigma_0,$
the decoherence functional (\ref{eq:threenine}) for the case of a 
pure initial state $\Psi$ may be written very simply in terms of
the branch wave functions (\ref{eq:branchwavefns}) as
\begin{eqnarray}
 D(h,h') &=& {\cal N}\ \sum_{i}\ p_{i}''\
\left(\Phi_{i}\circ\Psi_{h}\right)\left(\Phi_{i}\circ\Psi_{h'}\right)^{*}
         \nonumber\\
         &=&  {\cal N}\ \Psi_{h'}\circ\scri\circ\Psi_{h}.  \label{eq:Dscri}
\end{eqnarray}

In order to capture the notion of ``final indifference'', we next 
split $\Branch$ into orthogonal sectors $\Branch^{\pm}$ on which the 
Wheeler-DeWitt product $\circ$ is positive or negative definite.%
\footnote{%
We are not familiar with an explicit procedure for performing this split 
of the Bianchi IX minisuperspace in general. %
(See however \cite{ATU93,Marolf95,HM97,Embacher98}.)  
Nonetheless, as argued below, since we only need to evaluate the products
appearing in (\ref{eq:Dscri}) at very small volume, the arguments of Higuchi
and Wald \cite{HW95} noted above suggest that the construction we give when the
branch wave functions on $\sigma_0$ are in $\Fall(\sigma_{0})$ is sufficient
for our purposes.
} %
We require for final indifference that the $\Phi_i$ be chosen so that
\begin{equation}
    \scri \circ \psi^{\pm} = \pm\ \psi^{\pm},  
\label{eq:scrionpsipm}    
\end{equation}
where $\psi^{\pm} \in \Branch^{\pm}.$  

Such an identity $\scri$ may be explicitly constructed by choosing
bases $\{\Phi^{\pm}_i\}$ for $\Branch^{\pm}$ which are orthogonal in the inner 
product (\ref{eq:threesix}),
\begin{equation}
   \Phi^{\pm}_i \circ \Phi^{\pm}_j = \pm\ \delta_{ij}
\label{eq:finon}
\end{equation}
and
\begin{equation}
   \Phi^{\pm}_i \circ \Phi^{\mp}_j = 0.
\label{eq:finpmmpo}
\end{equation}
Setting all the $p_i''=1$, with such a basis we may construct projection 
operators
\begin{equation}
      \scri^{\pm}(q''_2,q''_1) 
         = \sum_{i}\ \Phi^{\pm}_{i}(q''_2)\Phi^{\pm*}_{i}(q''_1)
\label{eq:scripm}
\end{equation}
for which
\begin{equation}
     \scri^{\pm}(q''_2,q''_1) \circ \psi(q''_1) = \pm\ \psi^{\pm}(q''_2).  
\label{eq:scripmonpsi}
\end{equation}
$\scri$ is then given by
\begin{equation}
   \scri = \scri^+ + \scri^-
\label{eq:scridef}
\end{equation}
so that
\begin{equation}
     \scri \circ \psi = \psi^+ - \psi^-.  
\label{eq:scrionpsi}
\end{equation}

Explicit calculations require an explicit choice of the $\Phi^{\pm}_i.$ Since
we are restricting our attention to coarse grainings whose branch wave
functions $\Psi_{h}(q)$ lie in $\Fall(\sigma_{0}),$ we can implement the split
of the space of branch wave functions into $\Branch^{\pm}$ by choosing the $\{
\Phi_{i}^{\pm} \}$ to be of definite frequency on $\sigma_{0}$
with respect to $\alpha.$   That is, relabelling the index 
$i \rightarrow \vec{p}$, 
\begin{equation}
   \frac{\partial}{\partial\alpha}\ \Phi_{\vec{p}}^{\pm}(q)
    = \mp i\, \omega_{p}\, \Phi_{\vec{p}}^{\pm}(q).
                                            \label{eq:pmfreq}
\end{equation}
For this to be possible, it must be that the Wheeler-DeWitt
operator $H$ and $\partial/\partial\alpha$ commute.  In fact,
\begin{equation}
 \left[ \frac{\partial}{\partial\alpha},\, H \right]\, 
                           f(\alpha,\beta_{\pm},\phi)
   = 4e^{4\alpha}V_{\beta}(\beta_{\pm})\ f(\alpha,\beta_{\pm},\phi)
      + 6e^{6\alpha}  V_{\phi}(\phi) f(\alpha,\beta_{\pm},\phi),
                                      \label{eq:commute}
\end{equation} 
so that $\partial/\partial\alpha$ and $H$ approximately commute everywhere on
surfaces of large negative $\alpha$ so long as they act in the space of
functions $\Fall(\sigma_{0})$ 
for which $V_{\beta}f$ and $V_{\phi}f$ remain bounded.  Since $V_{\beta}$ and
$V_{\phi}$ are both bounded below, this is equivalent to the condition that $Vf
\approx 0$ on $\sigma_0$ -- or more precisely, that $e^{-4\alpha}Vf$ is bounded
-- whence our choice of boundary conditions on the allowed solutions to the
constraint in Section \ref{sec:df}.


The explicit form of the $\Phi^{\pm}_{\vec{p}}$ is not difficult to find. 
It is clear from (\ref{eq:twonine}) that on $\sigma_{0}$ there are
large regions near the origin in $\vec{q}$-space for which the potential $V$
is totally negligible so long as $V_{\phi}$ remains bounded.  
In fact, it is a standard part of the lore of mixmaster cosmologies%
\footnote{\cite{Misner70-72} 
are useful additions to the literature already cited.  The discussion in
section IV of \cite{HW95} provides an extremely useful formulation for general
type A Bianchi models; see also \cite{LW90} for Bianchi IX.  %
} %
that $V_a,$ the anisotropy potential for Bianchi IX -- see below
(\ref{eq:twoseven}) -- is well-approximated as $\alpha \rightarrow -\infty$ by
a triangular set of potential walls at $\alpha = -(\beta_+ \pm
\sqrt{3}\beta_-)$ and $\alpha = 2\beta_+$.  Well inside these walls, we may
take $V$ to be essentially zero for reasonable $V_{\phi}$.%
\footnote{Kucha\v{r} has shown \cite{Kuc81} that this ``asymptotic freedom'' 
of the gravitational potential $V$ is a generic feature of superspace, and is 
not special to homogeneous models.
} %
Near the origin of the spacelike surface $\sigma_{0}$, the
$\Phi_{\vec{p}}^{\pm}(q)$ can then be taken to have the 
form
\begin{equation}
  \Phi^{\pm}_{\vec{p}}(q) = \frac{1}{\sqrt{(2\pi)^n2\omega_{p}}}
               e^{\mp i\omega_{p}\alpha}e^{i\vec{p}\cdot\vec{q}},
\label{eq:asympform}
\end{equation}
where $\omega_p^2 = \vec{p}^2.$  %
In the asymptotic region where the potential $V$ is not negligible we may
choose to maintain {\it either} (\ref{eq:threefive}) or (\ref{eq:pmfreq}),
since our branch wave functions are by design essentially zero there.  Should
we choose to adhere strictly to the requirement (\ref{eq:threefive})
everywhere on $\sigma_0$, the $\Phi^{\pm}_{\vec{p}}$ will of course be more
complicated in the asymptotic region.
Practically, however, it is usually easiest to retain the form
(\ref{eq:asympform}) everywhere.  So long as we restrict attention to initial
states and coarse-grainings for which $\Branch|_{\sigma_0}\subset
\Fall(\sigma_0)$ the distinction will be practically irrelevant and the
explicit form (\ref{eq:asympform}) can be used in calculating the decoherence
functional.

The basis of functions (\ref{eq:asympform}) are
orthogonal in the appropriate sense,
\begin{equation}
\label{eq:phiON}
 \Phi_{\vec{p}}^{\pm} \circ \Phi_{\vec{p}'}^{\pm} = 
 \pm \delta^{(n)}_{\vec{p}\vec{p}'}, 
\end{equation} 
and 
\begin{equation}
  \Phi_{\vec{p}}^{\pm} \circ \Phi_{\vec{p}'}^{\mp} = 0. 
\end{equation} 
($n$ is the number of ``spatial'' minisuperspace degrees of freedom -- three, in
the present example, $\beta_{\pm}$ and $\phi$.)  The projections $\scri^{\pm}$
of (\ref{eq:scripm}) may then be explicitly constructed as 
(taking $\sum_i p_i''\rightarrow\int d^np$)
\begin{equation}
 \scri^{\pm}(q_2'',q_1'') = \int d^np\, 
 \Phi^{\pm}_{\vec{p}}(q_2'')\Phi^{\pm *}_{\vec{p}}(q_1'').
\label{eq:scripmexplicit}
\end{equation}    

Finally, we note that the orthonormal bases $\{\Phi^{\pm}_i\}$ are also useful
to represent the initial condition $\Psi.$ Indeed, 
solutions to the Wheeler-DeWitt equation in $\Fall(\sigma_o)$ may be written
as
\begin{eqnarray} 
\Psi &=& \sum_i c^+_i\Phi^+_i + \sum_i c^-_i\Phi^-_i \nonumber \\
     &=& \Psi^+ + \Psi^-
\end{eqnarray} 
where
\begin{equation}
c^{\pm}_i = \pm\, \Phi^{\pm}_i \circ \Psi.
\end{equation}
For the particular choice of basis (\ref{eq:asympform}) for $\Fall(\sigma_o)$, 
this looks like
\begin{equation}
  \Psi(q) = \int d^np\, 
  \left\{ \tilde{\Psi}^+(\vec{p})\Phi^+_{\vec{p}}(q) + 
         \tilde{\Psi}^-(\vec{p})\Phi^-_{\vec{p}}(q) \right\}
  \label{eq:initialdecomp}       
\end{equation}
where
\begin{equation}
 \tilde{\Psi}^{\pm}(\vec{p}) = \pm\, \Phi^{\pm}_{\vec{p}} \circ \Psi
\end{equation}
on $\sigma_0.$   %

\subsection{The Wheeler-DeWitt Product and the Decoherence Functional}
\label{sec:dffinddef}

With the resolution of the identity $\scri$ 
in hand, we define the positive definite Wheeler-DeWitt product $\bullet$ by

\begin{eqnarray}
\psi\bullet\psi &=& \psi \circ \scri \circ \psi  \nonumber\\
     &=& \psi^{+}\circ\psi^{+} - \psi^{-}\circ\psi^{-}. 
\label{eq:poskgprod}
\end{eqnarray}
In terms of this positive product,%
\footnote{Indeed, up to technical details, we expect the choices we have made
are in their effect 
equivalent to employing the ``Rieffel induced'' inner product to construct the
decoherence functional; see \cite{HalThor01,HM97,HalThor02} and references
therein for further discussion.  (\ref{eq:Dindiff}), employing the Rieffel
induced product, is thus a natural alternative definition of the decoherence
functional that automatically incorporates a notion of ``final indifference.''  
More generally, note that (\ref{eq:Dscri}) defines a positive product on
the space of branch wave functions.  When the $\{\Phi_i\}$ constitute a
complete set 
and all the $p_i''\neq 0$ this is a genuine non-degenerate inner product.  One
may thus regard the specification of ``final indifference'' as equivalent to
the problem of defining an inner product on the space of branch wave
functions.
} %
the decoherence functional may be written very simply as
\begin{equation}
D(h,h') = {\cal N}\ \Psi_{h'} \bullet \Psi_{h},    \label{eq:Dindiff}
\end{equation}
where ${\cal N}$ is $(\Psi_{u} \bullet \Psi_{u})^{-1}.$  
(\ref{eq:Dindiff}) is the form of the decoherence functional we will
use for the remainder of the discussion.  
The generalization of this formula to the case of an impure {\it set} 
of initial wave functions $\{ \Psi_{i}, p_{i}^{\prime} \}$ is obvious.

\subsection{Classical Initial Conditions}
\label{sec:initial}

Because we will be interested in initial conditions and coarse grainings which
distinguish classical behaviour, it is useful to discuss the conditions
necessary to specify classical solutions.

The classical equations of motion -- the reduced Einstein equations --
resulting from the variation $\delta S/\delta q = 0$ of the
Lagrangian action appearing in (\ref{eq:threeseventeena}) are,
in the proper time gauge (\ref{eq:threethirteen}),
\begin{equation}
\label{eq:eqnofmotion}
  \frac{1}{2N}\, \frac{d^2 q_A}{dt^2} = - N\ \nabla_{\kern-0.2em A} V,
\end{equation}
along with the Lagrangian version of the constraint
$\delta S/\delta N = 0:$
\begin{equation}
\label{eq:lconstraint}
  \frac{1}{4N}\, \frac{d q^A}{dt} \frac{d q_A}{dt} + N\, V =0.
\end{equation}
Fixing a classical solution starting on 
$\sigma_0$ therefore requires two pieces of minisuperspace data for each
$q^A$, subject to the constraint (\ref{eq:lconstraint}).  The initial data
cannot be chosen arbitrarily even if $N$ is regarded as an unknown to be
determined by (\ref{eq:lconstraint}): for arbitrary initial data there is no
guarantee that there exists a real 
constant $N$ that can satisfy the constraint
$\dot{\alpha}^2_0 = \dot{\vec{q}}^2_0 + 4N^2V(q_0).$
The initial data must be chosen so that $\dot{\alpha}^2_0 - \dot{\vec{q}}^2_0$
has the same sign as $V(q_0)$.
The $N$ thus fixed by (\ref{eq:lconstraint}) will then appear as a parameter
in the classical solution emanating from $\sigma_0$ specified by the initial
data $(q_0,\dot{q}_0).$  
(Of course, a reparametrization of the time $t$ that preserves the proper time 
gauge will yield a different value for $N$.)

\subsection{Quantum Initial States: Examples}
\label{sec:exampleinitial}

We choose our first example initial state $\Psi^{\mathit{WP}}$ to be a
positive frequency solution to (\ref{eq:threefive}) that is localised well
within mixmaster's potential walls on the initial surface, and also away from
regions of very large $V_{\phi}$.  Such an initial state will prefer a
particular classical path and predict approximately classical behaviour along
the corresponding classical solution.

$\Psi^{\mathit{WP}}$ by assumption solves (\ref{eq:threefive}) with $V \approx
0$ and is localized near some $\vec{q}_0$ on $\sigma_0$.  It may then be
represented as
\begin{eqnarray}
\Psi^{\mathit{WP}}(q) &=& \int d^3p\, 
       \tilde{\Psi}^{\mathit{WP}+}(\vec{p})\Phi^+_{\vec{p}}(q-q_0)  \nonumber \\
      &=& \int\frac{d^3p}{\sqrt{(2\pi)^3}}\frac{1}{\sqrt{2\omega_{p}}}
              e^{-i\omega_{p}(\alpha-\alpha_0)} 
              e^{i\vec{p}\cdot(\vec{q}-\vec{q}_0)} 
              \tilde{\Psi}^{\mathit{WP}+}(\vec{p})
\label{eq:locinitial}
\end{eqnarray}
for some $\tilde{\Psi}^{\mathit{WP}+}(\vec{p})$, where of course $\omega_p =
|\vec{p}|$.  Taking $\omega_p^{-1/2}\tilde{\Psi}^{\mathit{WP}+}(\vec{p})$ 
to be a Gaussian centered around some $\vec{p}_0$, for instance, yields a
$\Psi^{\mathit{WP}}(q)$ localized on $\sigma_0$ about $(\vec{q}_0,\vec{p}_0)$
to the greatest extent consistent with the uncertainty principle.

Alternatively, we could consider an initial state of WKB form,
\begin{equation}
\label{eq:WKBinitial}
\Psi^{\mathit{WKB}} (q) = A(q)\, e^{iW(q)} ,
\end{equation}
where $A(q)$ is in an appropriate sense slowly varying relative to $W(q).$  
Approximate calculations of the wave functions corresponding to the various
proposals for the initial condition of the universe tend to have components of
this form.  We also require that $A(q)$ be of compact support on $\sigma_0$ so
as to ensure that $\Psi_0 \in \Fall (\sigma_0),$ and that $W(q)$ be ``positive
frequency'' in the sense that $\partial W/\partial \alpha < 0$ 
on $\sigma_0.$ (This will turn out to correspond to an ``expanding'' initial
condition on the paths this wave function defines.)

We will examine both of these choices in the sequel.  First, however, we shall
move on to discuss the category of coarse grainings defining approximately
classical behaviour that we shall consider.

\section{Approximate Classicality}
\label{sec:semiclassical}

In this section, we apply the decoherence functional (\ref{eq:Dindiff}) for
$\Lambda = 0$, Bianchi IX cosmologies to coarse grainings which distinguish
between those paths in minisuperspace which behave (semi-) classically, and
those which do not.  We find that for suitable choices of the initial
condition $\Psi$, and for a suitable class of definitions of the semiclassical
coarse grainings, the universe is predicted to behave classically with
probability near one.  In particular, $\Lambda = 0$ Bianchi IX cosmologies are
predicted quantum-mechanically to recollapse just as they do classically
\cite{LW90}.  This is satisfying, not least because of the surprising
prediction of Wald's \cite{Wald83,HW95}\ rigorous canonical quantization of
Bianchi IX that closed quantum universes do not recollapse in a classical
fashion.%
\footnote{The trouble is rooted in the fact that Wald's quantization
essentially employs the volume of the universe (rather, $\alpha$) as a
``time'' variable in the canonical quantization procedure.  With this
parameterization there is no way to construct minisuperspace wave packets
which follow an approximately classical trajectory through both expanding and
recollapsing phases.  Moreover, there appears no sensible way to define an
operator momentum conjugate to $\alpha$ (which corresponds physically to the
expansion rate) that can have both positive and negative eigenvalues
\cite{Wald93}.  In other words, this quantization seems to have an ``arrow of
time" that precludes the construction of states which follow the full course
of a classical evolution.  Instead, time appears to ``freeze" as the universe
approaches its classical maximum size \cite{HW95}, in the sense that physical
variables cease to evolve.  %
}
(For another discussion of difficulties interpreting wave functions for 
classically recontracting cosmologies, see \cite{KieferZeh95}.)

We begin with a simple example to illustrate the general procedure we have in
mind, then go on in the subsequent sections to explore semiclassical
coarse-grainings and the corresponding class operators and branch wave
functions in greater detail.

\subsection{Coarse-Graining by a Single Trajectory}
\label{sec:singlepathcg}

\begin{figure}[!htbp]   
\includegraphics[scale=0.75]{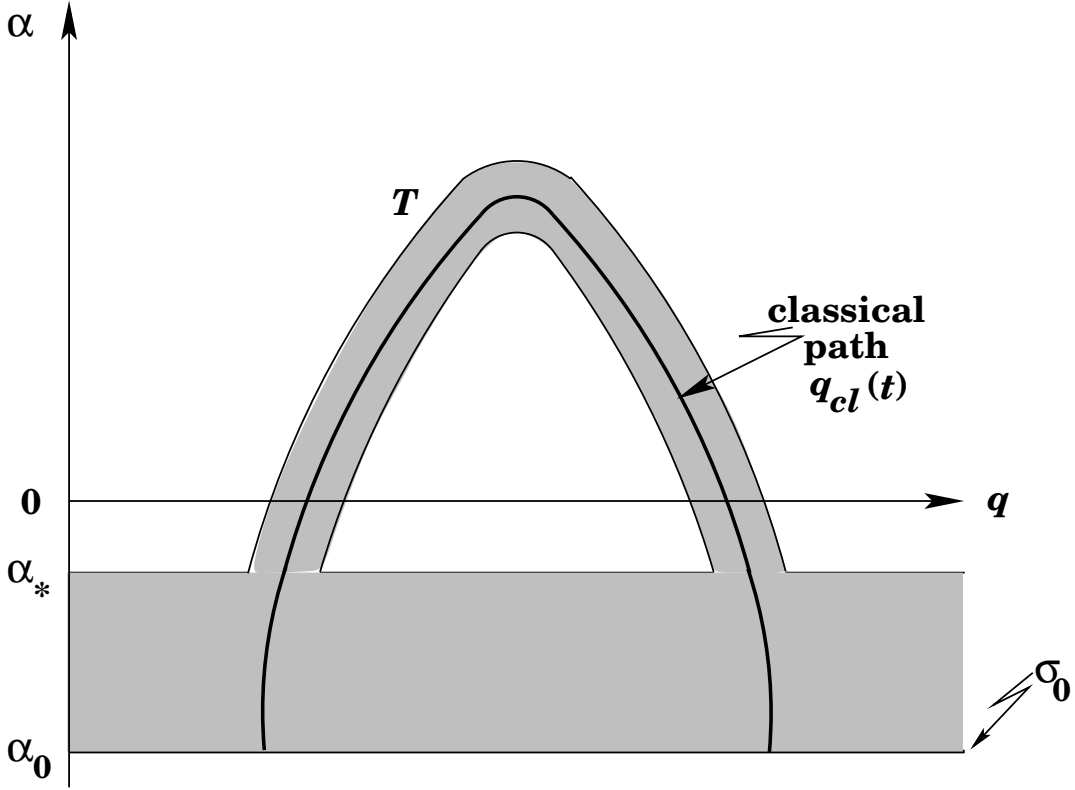}
\caption{\label{fig:tubecg} 
Schematic representation of a course-graining defining classical behaviour. 
The shaded region $T$ illustrated in the figure consists of all configuration
space for $\alpha < \alpha_*$, and for $\alpha > \alpha_*$, a tube surrounding
a trajectory $q_{cl}(t)$ that obeys the classical equations of motion.  The
fine grained cosmological histories can be divided into the class $c_T$ that
lie entirely inside $T$, and the class $c_{\overline{T}}$ that are sometimes
outside of it (possible only for $\alpha > \alpha_*$.)  The probability of
$c_T$ is the probability that the evolution of the universe is approximately
the classical evolution $q_{cl}(t)$ for $\alpha > \alpha_*$.  }
\end{figure}

As a particularly simple example of a semiclassical coarse-graining, ask
whether the model universe follows a {\em particular} trajectory in
minisuperspace $q_{cl}$.  %
To this end introduce the Euclidean distance on superspace
and define a region around the curve $q_{cl}$ -- say a tube $T$ of radius 
$\delta$.  (See Figure \ref{fig:tubecg}.%
\footnote{%
Note that strictly speaking, $T$ must not intersect $\sigma_0$ in order that
the coarse graining does not restrict paths on the initial surface.  This was
necessary in Section \ref{sec:df} to guarantee the independence of the
decoherence functional from the choice of initial surface.  We may choose
either to give up this independence for coarse-grainings of this kind -- not a
surprising or onerous restriction for coarse-grainings which explicitly
restrict paths on the initial surface -- or end the tube $T$ at a volume
$\alpha_*$ larger than $\alpha_0$, as in the figure.  
When $\alpha_*$ is close to $\alpha_0$,  %
or more generally for systems which are not chaotic,  %
we expect these approaches to be practially equivalent. 
})  %
We partition the paths $q(t)$ by whether they lie entirely in $T$ 
or not, labelling the corresponding classes $c_T$ and $c_{\overline{T}}$.

It is plausible that when $\delta$ is sufficiently large the path integral in
(\ref{eq:threeseventeenb}) can be done by the stationary phase approximation to
find
\begin{subequations}
\label{eq:tubeclassop}
\begin{eqnarray}
\Melt{q''}{C_T}{q'}  & \approx & \left\{
    \begin{array}{lcl}   
       \Delta(q'',q')e^{iS_{cl}(q'',q')} & &  q',q'' \in T  \\
        0                                & & \mathrm{else}
    \end{array}\right.                           \label{eq:tubeclassopa} \\
\Melt{q''}{C_{\overline{T}}}{q'}  & \approx & \left\{
    \begin{array}{lcl}   
       \Delta(q'',q')e^{iS_{cl}(q'',q')} & &  q'\ \mathrm{or}\ q'' \not\in T  \\
        0                                & &  q',q'' \in T
    \end{array}\right.                           \label{eq:tubeclassopb} 
\end{eqnarray}
\end{subequations}
so that $C_T + C_{\overline{T}} = C_u$ as required by (\ref{eq:threethree}),
where $C_u$ is the full, unrestricted propagator.   Here $\Delta$ is the usual
semiclassical prefactor and $S_{cl}$ is the action evaluated on the classical
path connecting $q'$ to $q''$.  (See Section \ref{sec:semiclassicalclassops}
for further details.)

\begin{figure}[!htbp]   
\includegraphics[scale=0.75]{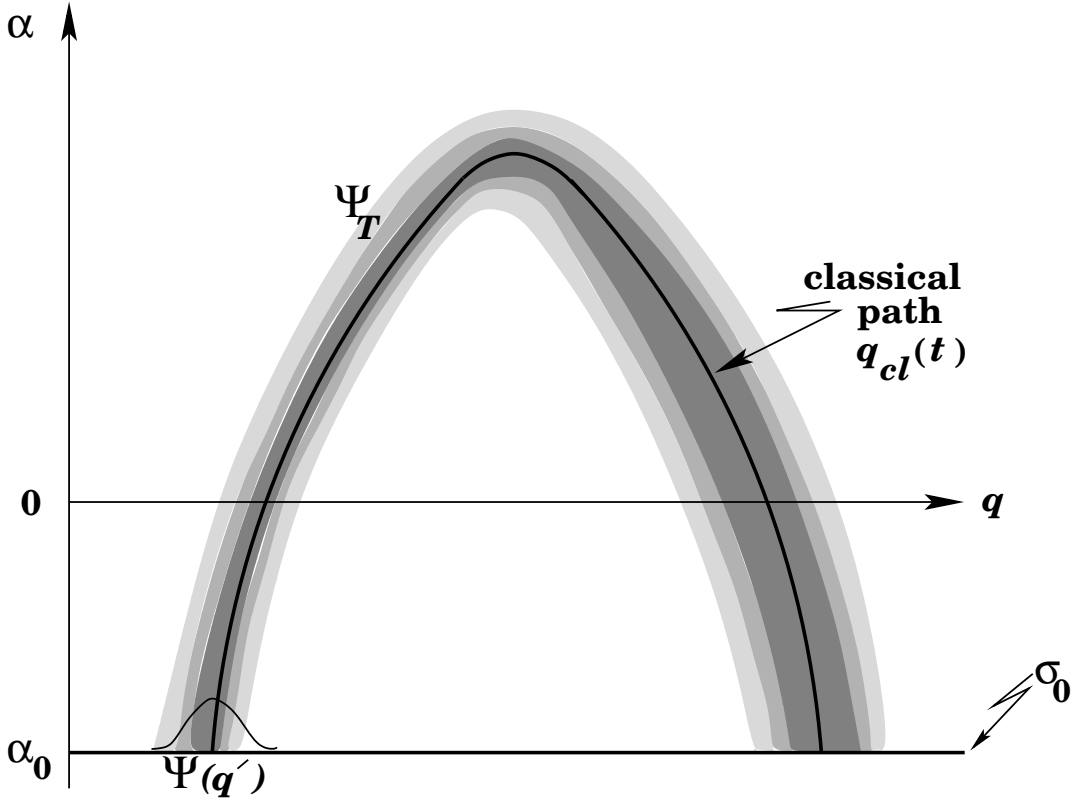}
\caption{\label{fig:Psi_T} 
Schematic representation of the region of support of the semiclassical branch
wave function $\Psi_T(q'')$ corresponding to the class of paths $c_T$ for a
localized initial condition $\Psi(q')$.  $\Psi_T(q'')$ is concentrated around
the classical trajectory $q_{cl}(t)$ preferred by the initial condition.
}
\end{figure}

Now consider the special case of an initial condition $\Psi(q')$ whose center
moves along $q_{cl}(t)$.  Given (\ref{eq:tubeclassop}),
\begin{subequations}
\label{eq:tubeBWF}
\begin{eqnarray}
\Psi_T(q'') & = & \Melt{q''}{C_T}{q'}\circ\Psi(q') \\
            & \approx & \left\{
    \begin{array}{lcl}   
       \Delta(q'',q')e^{iS_{cl}(q'',q')}\circ\Psi(q') & &  q'' \in T  \\
        0                                             & & \mathrm{else}
    \end{array}\right.                                 \label{eq:tubeBWFa} \\
\Psi_{\overline{T}}(q'') & \approx &   0               \label{eq:tubeBWFb}
\end{eqnarray}
\end{subequations}
(See Figure \ref{fig:Psi_T}.)
Thus we expect according to (\ref{eq:threeone}) and (\ref{eq:Dindiff}) the
probability for $\Psi_T$ to be approximately 1 and the probability for
$\Psi_{\overline{T}}$ to be approximately 0.  For these special choices of
coarse-graining and initial condition, therefore, the prediction is that the
model universe ``behaves semiclassically'' over the whole of its evolution 
from near the big bang to the big crunch.

\subsection{Coarse Grainings by Classical Paths}
\label{sec:classicalcg}

The coarse-graining of the previous section asks only whether the evolution of 
the universe is close to a {\it particular} $q_{cl}(t)$.  More generally we 
can ask for the probability that it evolves close to {\it any} classical path.
In this section we discuss the corresponding coarse-grainings.
 
Coarse-grainings that can serve to distinguish classical behaviour can be
characterized generally according to whether a minisuperspace trajectory is
``close to'' a classical trajectory in some appropriate sense.  Because a
classical 4-geometry is fixed by a solution $z(t) = (q^A(t),N)$ to the proper
time gauge Einstein equations (\ref{eq:eqnofmotion}) and
(\ref{eq:lconstraint}), we say that $z(t)$ is ``approximately classical" when
there exists a solution $z_{\mathit{cl}}(t)$ to Einstein's equations for which
\begin{equation}
    F[z(t),z_{cl}(t)] < 1,
\label{eq:distanceF}
\end{equation} 
where $F[\cdot,\cdot]$ is a reparametrization invariant distance
functional on the extended space of paths ${\cal C}_{ext}.$  
$F$ thereby partitions the space of paths into the class $c_{\mathit{cl}}$ of
those which behave classically, according to the standard $F$, and the class
which do not, $c_{\mathit{qm}}$.  An example of such a functional is the
Euclidean distance on superspace employed above, which can be rewritten in the
form (\ref{eq:distanceF}) simply by dividing by $\delta$.

More refined coarse grainings are also possible.  It is naturally of interest 
to know {\it to which} path(s) $z(t)$ may be close.  Various strategies are
possible; we describe one here that will be employed in the sequel.

First, partition the classical trajectories into classes $c_{ij}$, 
possibly by ranges $\{\Delta_i',\Delta_j''\}$ of initial and final position on 
$\sigma_0,$ or by ranges of initial position and momentum 
$\{\Delta_i',\Gamma_j'\}$.
The former might be of interest when we want to make predictions about
universes which, say, begin and end with low anisotropy; the latter, when we
are more interested in comparing universes with differing initial states.   
(Either way, the classical solutions provide a natural mapping  
$\{\Delta_i',\Delta_j''\} \leftrightarrow \{\Delta_i',\Gamma_j'\}$
for a given $\Delta_j''$ or $\Gamma_j'$.)
Next, for each quantum path, compute
\begin{equation}
\label{eq:minF}
  \inf_{z_{cl}(t)} F[z(t),z_{cl}(t)],
\end{equation}
say, by minimizing with respect to the initial conditions of the classical
paths.  
If this number is greater than one, then $z(t) \in c_{\mathit{qm}}.$ Otherwise,
collect $z(t)$ into the same class $c_{ij}$ as the $z_{cl}(t)$
which obtains the minimum.\footnote{For 
reasonable $F$'s, this should be a genuine partition of the paths.  In those
rare cases for which there are $z_{cl}(t)$ in different $c_{ij}$ which yield
the same minimum, additional criteria must be selected and applied to complete
the partition.  %
Similarly, if there are qualitatively distinct classical paths connecting
points in $\Delta_i'$ to $\Delta_j''$ -- in the case where there are multiple
classical paths connecting a single pair of points, for example -- then it may
under some circumstances be desirable to further partition the class $c_{ij}$. 
We shall take this as understood without attempting to incorporate it into our
notation explicitly.  %
} 

A particular choice of semiclassical coarse-graining entails the selection of
a particular distance functional $F$.  However, we expect reasonable
characterizations of semi-classical behaviour to display a certain robustness
and a consequent insensitivity to the details of the definition of the
coarse-graining (choice of $F$) for most practical applications. 
Nevertheless, we can make a few general observations about what constitutes an
acceptable classical distance functional on ${\cal C}_{ext}$.

First, $F$ must not be such as to imply that paths obeying (\ref{eq:distanceF}) 
possess finite action.  This is a consequence of the well known \cite{ColemanZJ}\ 
fact that paths of finite action contribute zero measure to functional
integrals.  Instead, it is the small but non-differentiable fluctuations about
the classical stationary points which dominate the integral.    %
In this sense, the coarse graining must not be ``too small'', or in other
words must not be so refined as to exclude the essential contributions of the
truly quantum paths in the neighborhood of the classical trajectories.  In
particular, coarse grainings which imply that paths obeying
(\ref{eq:distanceF}) are {\em differentiable} are excluded from consideration. 
(Since the potential $V$ is finite, 
the action will be finite if the kinetic terms are.)  %
This condition is necessary in order that we can make sense of the stationary
phase approximation, and in particular argue that the values of the functional
integrals are indeed well approximated by their stationary phase values
when integrating only over paths satisfying  (\ref{eq:distanceF}).

This stationary-phase condition may be satisfied by admitting as
semi-classical coarse grainings those which include all paths nearby each
classical trajectory with a minimum Euclidean distance scale set by
$1/\sqrt{\lambda}$, where the $\lambda$ are the eigenvalues of
$\delta^2 S[q_{cl}(t)]/\delta q^A(t')\, \delta q^B(t'')$.  %

Second, the coarse graining must not be ``too large'', in the sense that paths
in the class $c_{\mathit{cl}}$ of approximately classical paths must be
meaningfully distinct from those in the complementary class $c_{\mathit{qm}}.$
The criteria applied to make this distinction will in general be particular to
individual physical problems in which a notion of approximate classicality is
to be applied.  Nonetheless, for most semi-classical considerations, the scale
of such coarse-grainings will typically be far coarser than the lower
limits implied by the considerations of the previous paragraphs.  %

Finally, it is a generic feature of quantum mechanics that localized states
spread over time.  In the absence of a stabilization mechanism such as
environmental decoherence through interaction with other degrees of freedom in
a system \cite{DH92,GMH93,Giulini,Hal89,Pad89,HL90}, even a wave packet whose
center follows a classical trajectory may eventually spread sufficiently that
its behaviour is unrecognizably classical.  While such stabilization mechanisms
are widespread in the real universe (as opposed to toy models with few degrees
of freedom), the phenomenon of wave packet spreading may be a relevant factor
in determining whether a particular choice of coarse-graining is meaningfully
``semi-classical'' for a given physical problem.

There are, of course, many choices for the distance functional $F$ that will
satisfy the general requirements laid out above \cite{lesH,GMH93}.  However,
in accordance with the general expectation that predictions concerning
semiclassical behaviour will be relatively insensitive to the details of the
choice of $F$, the approximate calculations given in the sequel will be
sufficiently coarse that a detailed specification will not be needed.  When
necessary, we will typically have in mind something like the Euclidean
distance on superspace employed in Section \ref{sec:singlepathcg}.  (A more
careful assessment of suitable choices for the distance functional $F$ in the
context of ordinary quantum mechanics will be given elsewhere
\cite{dac04cg}.)

\subsection{Semiclassical Propagator}
\label{sec:semiclassicalclassops}

Let us now proceed with the approximate evaluation of the class operators for
the general category of semiclassical coarse-grainings described above.

In the $\dot{N}=0$ (proper time) gauge, the class operators 
[{\it cf.\ }(\ref{eq:threesixteen}-\ref{eq:threeseventeen})] read
\begin{equation}
\label{eq:ptclassops}
\Melt{q''}{C_h}{q'} = 
     \int_{c_h} dN_{\geq 0}\, \int_{c_h} \delta q_{q'}^{q''} \exp\left( 
  i \int_0^1 dt \left[ \frac{1}{4N}G_{AB}\dot{q}^A\dot{q}^B - NV \right] \right).
\end{equation}
The integral is over all paths from $q'$ to $q''$ in the class $c_h$.  
It is to be noted that, strictly speaking, the restricted propagator
$\Melt{q''}{C_h}{q'}$ need only be defined for $q'$ and $q''$ in $\sigma_0$. 
$c_h$, however, defines a (possibly empty) class of paths passing through any
$q''$ in minisuperspace, and when $q''$ is not in $\sigma_0$ the integral over
these paths is what is meant by the path integral in (\ref{eq:ptclassops}). 
The set of points in minisuperspace for which this class is non-empty will be
denoted $T_h$, which will therefore be the region of support in $q''$ of the
class operator (\ref{eq:ptclassops}) in minisuperspace.

Let us begin by approximately evaluating the full propagator, the case in
which $c_h$ in (\ref{eq:ptclassops}) happens to be the class of all paths
$c_u$ from $q'$ to $q''$.  A steepest descents approximation to the functional
integral over $q$ gives
\begin{equation}
\label{eq:semicu}
\Melt{q''}{C_u}{q'}  = 
   \int_0^{\infty} dN\, \Delta(q'',q';N)\, e^{ i S_{cl}(q'',q';N) }  ,
\end{equation}
where $S_{cl}(q'',q';N)$ is the action evaluated for the classical 
path $q_{cl}(q'',q':N)$ -- solution to (\ref{eq:eqnofmotion}) with 
the given value of $N$ -- connecting $q'$ to $q''$, and
\begin{eqnarray}
\Delta^{-2}(q'',q';N) &\propto& 
   \det\frac{\delta^2 S[q_{cl}(t)]}{\delta q^A(t')\, \delta q^B(t'')}
        \nonumber\\    
      &\propto&  \det\left[ 
        \frac{1}{2N} G_{AB}\frac{d^2}{dt^2} 
              + N\nabla_A \nabla_B V(q_{cl}) \right],  
\label{eq:det-2}
\end{eqnarray}
dropping a multiplicative factor of $\det[-i\delta(t-t')]$ in the last line
that will cancel in the decoherence functional.  
Similarly, neglecting%
\footnote{Heuristically, if we exponentiate the determinant via $\ln\, \det
{\cal O} = {\rm tr}\, \ln {\cal O}$, the $N$ dependence of the semiclassical
prefactor $\Delta(q'',q';N)$ is only logarithmic, as compared with that of
$S_{cl}(q'',q';N)$.  } the $N$ dependence of $\Delta(q'',q';N)$ in
(\ref{eq:semicu}), the stationary phase approximation picks out the ``zero
energy'' path(s) in (\ref{eq:semicu}) obeying the constraint
(\ref{eq:lconstraint}).
Thus,
\begin{equation}
\label{eq:semicucl}
\Melt{q''}{C_u}{q'}  \approx 
    \Delta(q'',q')\, e^{ i S_{cl}(q'',q') },
\end{equation}
where $N$ now has its classical dependence $N(q'',q')$ on the
endpoints of the path given by substituting $q_{cl}(q'',q';N)$
into the constraint (\ref{eq:lconstraint}).%
\footnote{Strictly speaking, it is the {\it time integral} of
(\ref{eq:lconstraint}) that is the constraint that emerges from the stationary
phase approximation to the integral over $N,$ {\it viz.\ }$
\frac{\partial}{\partial N} S[ q_{cl}(q'',q';N) ] = 0.$ However, the left hand
side of (\ref{eq:lconstraint}) is the ``energy integral'' of solutions to
$\delta S/\delta q = 0$ ({\it i.e.\ }(\ref{eq:eqnofmotion})), and
hence is actually constant, so that the integral sign may be dropped.}
When there is more than one classical path connecting $q'$ to $q''$, 
then there will be a sum over such paths in (\ref{eq:semicucl}).

According to the famous Van Vleck formula \cite{Schulman,Kleinert}, 
$\Delta(q'',q')$ may be evaluated as
\begin{equation}
\Delta(q'',q') =  \left( \frac{1}{2\pi i}
 \frac{\partial^2 S_{cl}(q'',q')}{\partial N^2_{cl}} \right)^{-\frac{1}{2}}
      \left( \frac{1}{2\pi i\hbar}\right)^{\frac{d}{2}}
      \left| \det
      \frac{\partial^2 S_{cl}(q'',q')}{\partial q''\partial q'} 
            \right|^{\frac{1}{2}} e^{-\frac{i\nu\pi}{2}}, 
\label{eq:vvdet}
\end{equation}
where $d=n+1$ is the number of minisuperspace degrees of freedom --
4, in the present example, $\alpha, \beta_{\pm},$ and $\phi$ -- and
$\nu$ is the ``Maslov index'' of the path $q_{cl}(t)$ 
\cite{Schulman,Kleinert,semiclass}.  Roughly speaking, 
it counts the number of negative eigenvalues of 
$\delta^2 S[q_{cl}(t)]/\delta q^A(t')\, \delta q^B(t'')$
along the trajectory $q_{cl}(t)$.
The first factor arises of course from the ordinary stationary
phase integral over $N.$

In subsequent sections we will go on to discuss the prediction of
semiclassical evolution of the universe for various choices of initial state. 
For this purpose it will be helpful to close this section with the observation
that, as a consequence of the choice of positive range for the lapse $N$ and 
the boundary conditions on the ends of the paths on $\sigma_0$,
$\Delta^*(q'',q')e^{ -i S_{cl}(q'',q') }$, considered as a function of
$\alpha'$, is positive frequency in $\alpha'$.  Recalling that the propagator
in (\ref{eq:threeeight}) or (\ref{eq:branchwavefns}) is not complex-conjugated
when forming the indicated products, this means that 
$\Delta(q'',q')e^{ i S_{cl}(q'',q') }$ overlaps only with the positive frequency 
part of the initial wave function $\Psi$.  $\Delta(q'',q')e^{ i S_{cl}(q'',q') }$ 
is similarly negative frequency in $\alpha''$, so that only the {\it negative}
frequency pieces of the final boundary conditions $\Phi_i$ contribute.
This is to be expected, as the full propagator given by the path integral 
(\ref{eq:threefourteen}) over all paths and with a positive range for the 
lapse is in essence ($i$ times) the Feynman propagator for this  
minisuperspace theory \cite{Hal88}.

To see why it is true that $\Delta^*(q'',q')e^{-iS_{cl}(q'',q')}$ is positive
frequency in $\alpha'$, we shall show that its Fourier transform only has
support for positive frequencies $\omega'$.  That is, consider
\begin{equation}
\label{eq:fourten}
\int d\alpha'\, e^{i\omega'\alpha'} \Delta^*(q'',q')e^{-iS_{cl}(q'',q')}.
\end{equation}
Again neglect the $\alpha'$ dependence of $\Delta^*(q'',q')$ as
logarithmically slower than that of $S_{cl}$.  The largest contribution to the
integral comes from the point of stationary phase
\begin{equation}
\label{eq:foureleven}
   \omega' = \frac{\partial S_{cl}}{\partial\alpha'}.
\end{equation}

Considering the action appearing in (\ref{eq:ptclassops}) as a functional of
classical paths, it is a straightforward matter to verify that under
variations of the endpoints of those classical paths
\begin{equation}
\label{eq:fourtwelve}
\delta S_{cl} = 
     \left. \frac{\dot{q}^A_{cl}}{2N_{cl}}\, \delta q_A  \right|^1_0
\end{equation}
where $q^A_{cl}(t)$ is the classical path joining 
$q' = q_{cl}(0)$ to $q''=q_{cl}(1)$.  Thus
\begin{equation}
\label{eq:fourthirteen}
\frac{\partial S_{cl}}{\partial\alpha'} =
    \frac{\dot{\alpha}_{cl}(0)}{2N_{cl}}.
\end{equation}

As we have chosen $N$ to be positive, and $\dot{\alpha}_{cl}(0)$ is positive
for all classical paths%
\footnote{More generally, recall that in Section \ref{sec:fighs} we chose
``expanding'' initial and ``contracting'' final boundary conditions for {\it
all} paths, not just the classical ones, {\it i.e.\ }we required
$\dot{\alpha}(0)>0$ and $\dot{\alpha}(1)<0$ for all paths.
} %
that begin on our boundary surface of large negative $\alpha$, we see that
only a positive $\omega'$ contributes significantly to (\ref{eq:fourten}), so
that $\Delta^*(q'',q')e^{-iS_{cl}(q'',q')}$ is positive frequency in
$\alpha'$.  Similarly $\Delta(q'',q')e^{+iS_{cl}(q'',q')}$ is {\it negative}
frequency in $\alpha''$.  This means that
$\Delta(q'',q')e^{+iS_{cl}(q'',q')}$ is negative frequency in {\it both}
$\alpha'$ and $\alpha''$.  (This is a consequence of the fact that the
classical paths both begin and end on the same surface, so that
$\dot{\alpha}_{cl}(0) > 0$ while $\dot{\alpha}_{cl}(1) < 0$ on $\sigma_0$. 
This difference cancels the sign difference between the endpoints that arises
in the variation of the action (\ref{eq:fourtwelve}).)  As noted, the
semiclassical class operator $\Delta(q'',q')e^{+iS_{cl}(q'',q')}$ therefore
couples only to the positive frequency components of the initial state
in the branch wave functions (\ref{eq:branchwavefns}).
The branch wave functions in turn will be negative frequency at the endpoints
of the paths on $\sigma_0$.

\subsection{Semiclassical Class Operators}
\label{sec:cgsemiclassops}

Now consider the class operators $\Melt{q''}{C_h}{q'}$ corresponding to the
semiclassical coarse-grainings discussed in Section \ref{sec:classicalcg}. 
For the simplest partition of the fine-grained paths into the class
$c_{\mathit{cl}}$ of those which behave classically according to
(\ref{eq:distanceF}), and those which do not, $c_{\mathit{qm}}$, it is clear
from the preceeding section that since the leading contribution to
(\ref{eq:ptclassops}) comes from the classical path connecting $q'$ to $q''$,
to leading order
in the stationary phase approximation
\begin{subequations}
\label{eq:semiclclassop}
\begin{eqnarray}
\Melt{q''}{C_{cl}}{q'} & \approx & \Delta(q'',q') e^{ iS_{cl}(q'',q') } 
                                               \label{eq:semiclclassopa} \\
                       & \approx & \Melt{q''}{C_u}{q'} 
                                               \label{eq:semiclclassopb}
\end{eqnarray}
\end{subequations}
while%
\begin{equation}
\label{eq:qmclassop}
\Melt{q''}{C_{qm}}{q'} \approx 0.
\end{equation}
(The zero on the right hand side of (\ref{eq:qmclassop}) and all such
subsequent formul\ae\ need only be true in the sense of a distribution, {\it
e.g.\ }in the case in which the propagator is such a rapidly oscillating
function that the overlap integral in (\ref{eq:branchwavefns}) is 0 for any
initial state.)

For the more refined partition defined in Section \ref{sec:classicalcg} of
$c_{\mathit{cl}}$ into sub-classes $c_{ij}$ of semiclassical paths according
to ranges
$\{\Delta_i',\Delta_j''\} \leftrightarrow \{\Delta_i',\Gamma_j'\}$  %
of initial and final positions on $\sigma_0$ or the corresponding initial
positions and momenta,
\begin{equation}
\label{eq:semiclclassopij}
\Melt{q''}{C_{ij}}{q'}   \approx  \left\{
    \begin{array}{lcl}   
       \Melt{q''}{C_{cl}}{q'} & &  q' \in \Delta_i', \ q'' \in T_{ij}  \\
        0                             & & \mathrm{else}
    \end{array}\right.                           
\end{equation}
Here $T_{ij}$ is the region of support in $q''$ of the class operator 
$\Melt{q''}{C_{cl}}{q'}$, as defined following (\ref{eq:ptclassops}).
(See Figure \ref{fig:Psi_ij}.)

These semiclassical results for the class operators are useful only to the
extent that the stationary phase approximation can be trusted.  Some
techniques for evaluating restricted functional integrals such as
(\ref{eq:ptclassops}) are discussed in \cite{lesH}.  Further strategies for
more detailed calculation of such integrals in the context of ordinary quantum
mechanics are under investigation \cite{dacyue04}.

\subsection{Semiclassical Branch Wave Functions}
\label{sec:cgsemiBWFs}

Given the semiclassical class operators (\ref{eq:semiclclassop} -
\ref{eq:qmclassop}), the corresponding branch wave functions
(\ref{eq:branchwavefns}) can be evaluated as follows. 
Consider an initial state $\Psi(q')$.  Then
\begin{subequations}
\label{eq:semiclBWF}
\begin{eqnarray}
\Psi_{\mathit{cl}}(q'') 
    & = &  \langle q'' \| C_{\mathit{cl}} | \Psi \rangle   
                                                     \label{eq:cgsemiBWFsa} \\
    & = &  \Melt{q''}{C_{\mathit{cl}}}{q'}\circ\Psi(q') \label{eq:cgsemiBWFsb} \\
    & \approx & \langle q'' \| C_u | \Psi \rangle
                                                     \label{eq:cgsemiBWFsc} 
\end{eqnarray}
\end{subequations}
{\it i.e.\ }$\Psi_{\mathit{cl}}(q'')$ is approximately simply the initial state
propagated forward by the full propagator, while
\begin{equation}
\label{eq:qmBWF}
\Psi_{\mathit{qm}}(q'') \approx 0.
\end{equation}
Note that (\ref{eq:semiclBWF}-\ref{eq:qmBWF}) are independent of any
particular properties of the initial state $\Psi(q')$, and are wholly a
consequence of (\ref{eq:semiclclassop} - \ref{eq:qmclassop}).

Now consider the more refined coarse-graining of $c_{\mathit{cl}}$ into
sub-classes $c_{ij}$ of approximately classical paths.  Denote the regions of
position and momentum support of $\Psi(q')$ on $\sigma_0$ by $\varepsilon$ and
$\gamma$.  Then
\begin{subequations}
\label{eq:semiclBWFij}
\begin{eqnarray}
\Psi_{ij}(q'') & = &  
        \Melt{q''}{C_{ij}}{q'}\circ\Psi(q') \label{eq:semiclBWFija} \\
               & \approx & \left\{
    \begin{array}{lcl}   
                     & & \epsilon \cap \Delta_i' \neq \emptyset, \\              
          \langle q'' \| C_{\mathit{cl}} | \Psi \rangle                                                 
                                       & &  \gamma \cap \Gamma_j' \neq \emptyset,  \\
                                       & &  q'' \in T_{ij}  \\
        0                             & & \mathrm{else}
    \end{array}\right.   \label{eq:semiclBWFijb} 
\end{eqnarray}
\end{subequations}
with
\begin{equation}
\label{eq:semiclBWFsum}
\Psi_{\mathit{cl}}(q'') = \sum_{ij} \Psi_{ij}(q'').
\end{equation}

\begin{figure}[!htbp]   
\includegraphics[scale=0.75]{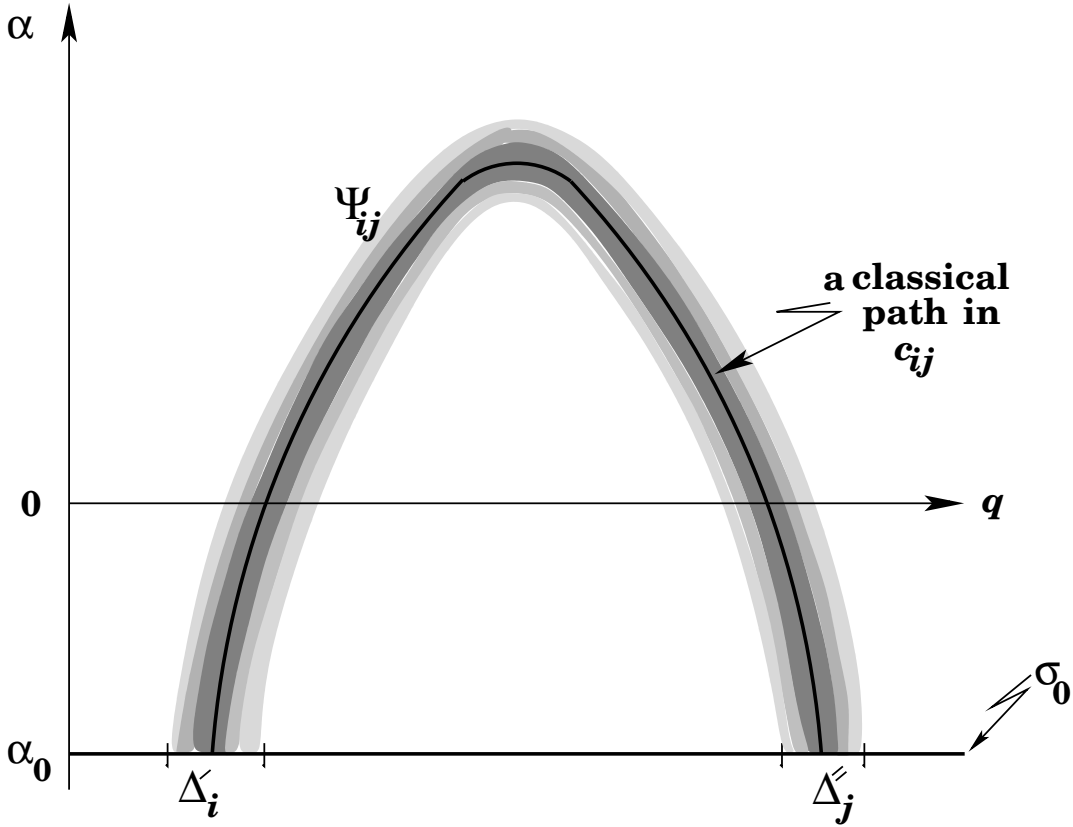}
\caption{\label{fig:Psi_ij}   
Schematic representation of the region of support of the semiclassical branch
wave function $\Psi_{ij}(q'')$ corresponding to the class of paths $c_{ij}$.
$\Psi_{ij}(q'')$ is concentrated around the classical trajectories connecting
$\Delta_i'$ to $\Delta_j''$, a representative of which is plotted in the
figure.  The contributions of each path are weighted by the value of the
initial condition $\Psi(q')$ in $\Delta_i'$.  $\Psi_{ij}(q'')$ is positive
frequency in $\Delta_i'$ and negative frequency in $\Delta_j''$.  Assuming
$\Psi(q')$ is not zero in $\Delta_i'$, the shaded region will be roughly the
same as $T_{ij}$, the region of support of the class operator
(\ref{eq:semiclclassopij}) on minisuperspace.
}%
\end{figure}

Note that for this class of coarse-grainings, given (\ref{eq:semiclBWF}) and
(\ref{eq:semiclBWFij}) we may expect the branch wave functions to
approximately satisfy the constraint (\ref{eq:threefive}), even though the
coarse-graining does restrict the paths on the initial surface (see Section
\ref{sec:df}.)  Indeed, according to (\ref{eq:semiclBWFijb}), $\Psi_{ij}(q'')$
is just the 
initial data $\{\Delta_i',\Gamma_j'\}$ selected by the coarse-graining and
weighted by the initial state $\Psi(q')$, %
propagated by the Wheeler-DeWitt propagator (\ref{eq:semiclclassop}).  %
As a function on minisuperspace it is therefore concentrated around the
classical trajectories connecting $\Delta_i'$ to $\Delta_j''$.  (See Figure
\ref{fig:Psi_ij}.)
On $\sigma_0$, when $q'' \in \Delta_i'$ $\Psi_{ij}(q'')$ is essentially the
part of the initial condition with momenta in $\Gamma_j'$.  In other words, 
\begin{equation}
  \Psi_{ij}(q'') \approx \int\nolimits_{\Gamma_j'} \hspace{-5pt} d^np\, 
              \tilde{\Psi}^+(\vec{p})\Phi^+_{\vec{p}}(q'')  
    \label{eq:Psi_ijDeltai}
\end{equation}
on $\sigma_0$ when $q'' \in \Delta_i'$, and is approximately zero otherwise.
The restriction to positive frequency is a consequence of the fact -- see
Section \ref{sec:semiclassicalclassops} -- that the semiclassical propagator
(\ref{eq:semicucl}) couples only to the positive frequency part of the initial
condition.  Similarly, the semiclassical propagator is negative frequency in
$\alpha''$ at the endpoints of the paths on $\sigma_0$, so that
$\Psi_{ij}(q'')$ is negative frequency when $q'' \in \Delta_j''$.  We will put
these observations to use in the next section.

\subsection{Evaluation of the Decoherence Functional: 
Approximately Classical Behaviour}
\label{sec:semiclassicalpredictions}

Given the branch wave functions (\ref{eq:semiclBWF}-\ref{eq:semiclBWFij}),
evaluation of the decoherence functional (\ref{eq:Dindiff}) for the partitions
into approximately classical {\em vs.\ }highly quantum paths we have 
considered is straightforward.

For the coarsest-graining of paths into the classes $c_{\mathit{cl}}$ -- 
paths which are close to {\it some} classical path -- or the remainder
$c_{\mathit{qm}}$, decoherence is automatic because there is only one 
non-trivial branch wave function (\ref{eq:semiclBWF}).  Hence
\begin{eqnarray}
    p_{\mathit{cl}} & = & D(\mathit{cl},\mathit{cl}) \nonumber \\
     & \approx & 1, 
     \label{eq:probclassical}
\end{eqnarray}
{\it i.e.\ }the universe is predicted to approximately follow {\it some}
classical trajectory -- though we cannot say which one without a more refined
coarse-graining -- while
\begin{eqnarray}
    p_{\mathit{qm}} & = & D(\mathit{qm},\mathit{qm}) \nonumber \\
     & \approx & 0 \label{eq:probquantum}
\end{eqnarray}
independent of the choice of initial state on account of (\ref{eq:qmclassop}).%
\footnote{%
While this would seem to be a strong statement, it is more a reflection (via
(\ref{eq:semiclclassop}-\ref{eq:qmclassop})) of the fact that path integrals
are dominated by paths close to the classical path than it is a claim about
the actual classicality of the universe's behaviour.  One would hardly expect
a universe to behave in an approximately classical fashion under all
circumstances or for all choices of initial state.  Indeed, whether a state
behaves in a recognizably classical fashion is in general as much a property
of the initial state as it is of the characteristics of the restricted
propagator.  In this regard it is thus wise not to over-interpret the physical
significance of the prediction as it stands, in that initial states normally
thought of as highly non-classical such as {\it e.g.\ }``Schr\"{o}dinger's
Cat'' states -- superpositions of distinct localized wave packets -- lead to a
single non-trivial branch wave function $\Psi_{\mathit{cl}}$ just like any
other initial state ({\it cf.\ }(\ref{eq:semiclBWFsum}).)  To detect the
classically unusual features of such states {\it requires a more refined
coarse-graining} such as the partition of $c_{\mathit{cl}}$ into the classes
$c_{ij}$.  A fuller understanding of this critical issue requires a more
careful evaluation of the coarse-grained class operators and corresponding
branch wave functions than we can give here \cite{dac04cg}.
} %

To ask the question of {\it which} trajectory the universe follows for a given
choice of initial state requires the more refined partition of
$c_{\mathit{cl}}$ into the classes $c_{ij}$.  Because the branch wave
functions, and hence the decoherence functional (\ref{eq:Dindiff}), depend in
this case more sensitively on the choice of initial state, we will discuss the
examples of a localized initial state $\Psi_0$ and an initial state of WKB
form $\Psi^{\mathit{WKB}}$ in their turn.

\subsubsection*{Initial State of Localized Form}

A localized choice of initial state such as (\ref{eq:locinitial}), peaked about
a particular set of classical initial conditions $(\vec{q}_0,\vec{p}_0)$, leads
to a decoherence functional which makes the semiclassical prediction that the
universe (approximately) follows the classical path determined by the initial
conditions $(\vec{q}_0,\vec{p}_0)$ as follows.  In general, for a localized
wave packet $\Psi^{\mathit{WP}}(q')$, define as above the regions of
$\Psi^{\mathit{WP}}$'s position and momentum support on $\sigma_0$ by
$\varepsilon$ and $\gamma$.  Take $\vec{q}_0 \in \Delta_I'$ and $\vec{p}_0 \in
\Gamma_J'$ for some $(I,J)$, and assume for simplicity that $\varepsilon$ is
largely contained in $\Delta_I'$ and $\gamma$ is primarily concentrated in
$\Gamma_J'$.  (Because of the uncertainty principle, $\varepsilon$ and $\gamma$
cannot both be compact.)  In this case, through (\ref{eq:semiclBWFij}) there
will then be only one nontrivial branch wave function
$\Psi^{\mathit{WP}}_{IJ}(q'')$ and decoherence will again be automatic, with
the model universe consequently predicted with high probability to behave
approximately  like 
the classical trajectories in the class $c_{IJ}$.

To see explicitly how this comes about, consider
\begin{eqnarray}
 \Psi^{\mathit{WP}}_{cl}(q'') &\equiv& 
                 \langle q'' \| C_{cl} | \Psi^{\mathit{WP}} \rangle
                       \nonumber \\ 
        &\equiv& \Melt{q''}{C_{cl}}{q'} \circ \Psi^{\mathit{WP}}(q') 
                       \nonumber \\
        &\approx&  
       i \int\nolimits_{\sigma_0} d^3q'\, \Delta(q'',q')e^{iS_{cl}(q'',q')}
 \frac{\buildrel\leftrightarrow \over\partial}{\partial\alpha'} 
               \Psi^{\mathit{WP}}(q').
\label{eq:psiloccl}
\end{eqnarray}
For $\Psi^{\mathit{WP}}(q)$ localized around $\vec{q}_0$, it is clear that
only paths with initial points $\vec{q}'$ near $\vec{q}_0$ contribute
significantly.  Similarly, inserting (\ref{eq:locinitial}) and dropping %
a term containing $\partial \Delta(q'',q')/\partial \alpha'$,  
\begin{equation}
\Psi^{\mathit{WP}}_{cl}(q'') \approx
    \int\nolimits_{\sigma_0} \kern-.3em d^3q' 
    \int\kern-.3em\frac{d^3p}{\sqrt{(2\pi)^3}}\frac{1}{\sqrt{2\omega_{p}}}    
      \left\{ \omega_p + \frac{\partial S_{cl}(q'',q')}{\partial\alpha'}\right\}
      \Delta(q'',q')e^{iS_{cl}(q'',q')}   e^{i\vec{p}\cdot(\vec{q}'-\vec{q}_0)} 
      \tilde{\Psi}^{\mathit{WP}+}(\vec{p}).
\label{eq:psilocclsp}
\end{equation}
Again neglecting the $q'$ dependence of $\Delta(q'',q')$ as logarithmically
suppressed, in the stationary phase approximation the dominant contribution to
the $\vec{q}'$ integral comes when
\begin{equation}
\label{eq:psemicl}
   \vec{p} \approx  \left.
  - \frac{\partial S_{cl}(q'',q')}{\partial\vec{q}'} \right|_{q'=q_0}.
\end{equation}
The right hand side of this equation is the initial momentum
of the classical path connecting $q'$ to $q''$.%
\footnote{%
Note that the specification of $\dot{\alpha}(0)$ is implicit. 
To see this, note that $\partial S_{cl}/\partial \alpha$ is not independent of
$\partial S_{cl}/\partial \vec{q}:$ these quantities are related by the 
Hamilton-Jacobi equation %
$\nabla_{\kern-0.2em A} S\nabla^A S + V =0$ %
for $S_{cl}$ -- effectively the constraint (\ref{eq:lconstraint}) with 
$p_A = \dot{q}_A/2N = -\nabla_{\kern-0.2em A} S$.  
Therefore when (\ref{eq:psemicl}) holds, $\dot{\alpha}$ is given by
$\partial S/\partial \alpha$ through (\ref{eq:fourthirteen}); $N$ is
determined by demanding consistency of the solution with the constraint ({\it
cf.\ }the variant of this argument we give below while discussing the case of
an initial state of WKB form.)
} %
%
%
Since our initial wave functions only have support on regions of $\sigma_0$ 
in which $V \approx 0$, the Hamilton Jacobi equation for $S_{cl}$ implies that
$(\partial S_{cl}/\partial\vec{q})^2 = (\partial S_{cl}/\partial\alpha)^2$.  
From (\ref{eq:fourthirteen}), $\partial S_{cl}/\partial\alpha'$ is positive,
so that when (\ref{eq:psemicl}) holds,
$\omega_p = p \approx \partial S_{cl}/\partial\alpha'$, and thus
\begin{equation}
\Psi^{\mathit{WP}}_{cl}(q'') \approx
    \int\frac{d^3p}{\sqrt{(2\pi)^3}} \sqrt{2\omega_{p}}\, 
       \tilde{\Psi}^{\mathit{WP}+}(\vec{p}) 
    \int\nolimits_{\sigma_0} d^3q'\,  
       \Delta(q'',q')e^{iS_{cl}(q'',q')} e^{i\vec{p}\cdot(\vec{q}'-\vec{q}_0)}.                 
\label{eq:psilocclspsp2}
\end{equation}
Performing the integration over $\vec{p},$ we therefore see that 
the branch wave function is suppressed when
$\tilde{\Psi}^{\mathit{WP}}(\vec{p})$ is small in the neighborhood of the
initial momentum of the classical path connecting $q_0$ to $q''.$ The choice
of initial condition $\Psi^{\mathit{WP}}(q')$ on $\sigma_0$, therefore,
singles out the classical path specified by $(\vec{q}_0,\vec{p}_0)$: %
$\Psi^{\mathit{WP}}_{cl}(q'')$ is peaked around the classical path preferred
by the initial condition.
Through (\ref{eq:semiclBWFij}) it is then clear that the only nontrivial
branch wave function for this coarse-graining is the
$\Psi^{\mathit{WP}}_{IJ}(q'')$ corresponding to the class $c_{IJ}$ which 
contains the classical path determined by the initial conditions 
$(\vec{q}_0,\vec{p}_0)$:
\begin{equation}
\Psi^{\mathit{WP}}_{ij}(q'') \approx  \left\{
    \begin{array}{lcl}   
          \Psi^{\mathit{WP}}_{cl}(q'') & &  i=I,\ j=J  \\
        0                             & & \mathrm{else}
    \end{array}\right.    \label{eq:semiclBWFIJ}
\end{equation}
This is merely a more refined statement of (\ref{eq:tubeBWF}).

\subsubsection*{Initial State of WKB Form}

The situation for the WKB choice of initial state (\ref{eq:WKBinitial})
is a little different, though familiar from conventional treatments 
of quantum cosmology.

Let us evaluate the classical branch wave function corresponding 
to an initial state of the WKB form (\ref{eq:WKBinitial}).
\begin{subequations}
\begin{eqnarray}
\label{eq:WKBbwf}
\Psi^{\mathit{WKB}}_{cl}(q'') 
               &=& \Melt{q''}{C_{\mathit{cl}}}{q'} \circ \Psi^{\mathit{WKB}}(q')  
                                 \label{eq:WKBbwfa}\\
%
%
    &=& i \hspace{-3pt} \int\nolimits_{\sigma_0} \hspace{-9pt} d \Sigma'^A\  \hspace{-7pt}
    \left\{ \hspace{-3pt} A(q')  \hspace{-4pt}
    \left[ \hspace{-2pt} \Melt{q''}{C_{\mathit{cl}}}{q'} \hspace{-3pt}
         \buildrel{\kern-0.17em\leftrightarrow}\over{\nabla'}_{\kern-0.35em A} 
       e^{iW(q')} \right]   \hspace{-4pt} -  \hspace{-2pt}
         \Melt{q''}{C_h}{q'} e^{iW(q')} \nabla_{\kern-0.2em A}' A(q') 
             \hspace{-3pt} \right\}
                                  \label{eq:WKBbwfd}  \\
%
%
         &\approx&  
      i \hspace{-3pt} \int\nolimits_{\sigma_0} \hspace{-7pt} d \Sigma'^A\ 
          \hspace{-3pt} \Delta(q'',q')\, e^{i S_{cl}(q'',q')}\, A(q')\,  e^{iW(q')} 
          \hspace{-2pt} \left[ \nabla'_{\kern-0.2em A} W(q') - 
          \nabla'_{\kern-0.2em A} S_{cl}(q'',q')  \right] \hspace{-2pt}.
\label{eq:WKBbwfe}
\end{eqnarray}
\end{subequations}
In (\ref{eq:WKBbwfe}) we have dropped the terms containing
gradients of the more slowly varying terms $A(q')$ and
$\Delta(q'',q').$  In this form, it is clear that the dominant
contribution to the integral comes when the stationary phase condition
\begin{equation}
\label{eq:WKBmom}
  \nabla'_{\kern-0.2em A} W(q') = - \nabla'_{\kern-0.2em A} S_{cl}(q'',q').
\end{equation}
is satisfied.  The result is thus
\begin{equation}
\Psi^{\mathit{WKB}}_{cl}(q'') \approx
      2i\int\nolimits_{\sigma_0} d \Sigma'^A\ 
          \Delta(q'',q')\, A(q')\, \nabla'_{\kern-0.2em A} W(q')\, 
          e^{i S_{cl}(q'',q')}\,  e^{iW(q')}. 
\label{eq:psiWKBcl}
\end{equation}

The quantity on the right hand side of (\ref{eq:WKBmom}) is just the initial
momentum of the classical path connecting $q'$ to $q''.$ Note that it
is a statement about the complete minisuperspace
gradient of $S_{cl},$ not merely its projection into %
the $\alpha = {\mathit constant}$ surface %
$\sigma_0.$ We are free to assert that contributions to the integral
(\ref{eq:WKBbwfe}) are suppressed when (\ref{eq:WKBmom}) does not hold because
of the independence of the branch wave functions on the choice of initial
surface: wiggling $\sigma_0$ around a bit in the neighborhood of $q',$ yet
keeping $q'$ fixed on $\sigma_0,$ yields a family of stationarity conditions
for the dominant contribution to the branch wave functions.  For them all to
be true, (\ref{eq:WKBmom}) must hold.  Alternatively, note that
$\Psi^{\mathit{WKB}}(q)$ is a solution to the constraint (\ref{eq:threefive}). 
Dropping gradients of $A(q)$ at lowest order, this means that $W(q)$ obeys the
same Hamilton-Jacobi equation as does $S_{cl}(q'',q')$.  Thus, with
``expanding'' initial conditions on the paths, this means that the ``time''
component of (\ref{eq:WKBmom}) holds whenever the ``spatial'' part does, so
that (\ref{eq:WKBmom}) follows from its projection into $\sigma_0.$

The significance of (\ref{eq:WKBmom}) is that $\Psi^{\mathit{WKB}}_{cl}(q'')$ is
peaked around the {\it family} of classical trajectories emanating from points
$q'_0$ on $\sigma_0$ which have initial momentum $\nabla'_A W(q')|_{q'=q'_0},$
with a relative weight controlled by the WKB prefactor $A(q').$ This is simply
because $\Psi^{\mathit{WKB}}_{cl}(q'')$ is suppressed when the classical path
from $q'$ to $q''$ does not have initial momentum given by (\ref{eq:WKBmom}). 
When the classical trajectory from $q'$ to $q''$ has momentum $\nabla'_A W(q'),$
but $A(q')$ is small, the integral is similarly suppressed.


Now let us consider the situation of the more refined coarse-graining
$(c_{\mathit{qm}},c_{ij})$, where the classes $c_{ij}$ defined in Section
\ref{sec:classicalcg} are the partitions of semiclassical paths according to
ranges $\{\Delta_i',\Delta_j''\} \leftrightarrow \{\Delta_i',\Gamma_j'\}$ of
initial and final positions on $\sigma_0$ or the corresponding initial
positions and momenta.  Clearly, only the classes $c_{ij}$ containing paths
with initial points $q'$ in regions where $A(q')$ is not small can have
non-negligible branch wave functions.  Of these, the classes containing the
classical path emanating from $q'$ with initial momentum $\nabla'_A W(q')$
will possess the dominant branch wave functions; the {\it relative} magnitude
of these branch wave functions will be controlled by the WKB amplitude $A(q).$
These wave functions will as before be concentrated around the corresponding
classical trajectories, and are given as in (\ref{eq:semiclBWFij}) by
\begin{equation}
\Psi^{\mathit{WKB}}_{ij}(q'') \approx \left\{
\begin{array}{lcl}   
                     & & \epsilon \cap \Delta_i' \neq \emptyset, \\              
          \Psi^{\mathit{WKB}}_{cl}(q'')                                                 
                                       & &  \gamma \cap \Gamma_j' \neq \emptyset,  \\
                                       & &  q'' \in T_{ij}  \\
        0                             & & \mathrm{else}
    \end{array}\right.   
\label{eq:semiclWKBBWFij}
\end{equation}

For these more refined coarse-grainings, therefore, we have a {\it collection}
of non-trivial branch wave functions, and the decoherence functional has more
to tell us.  It is in particular of interest to ask whether the
coarse-graining $(c_{\mathit{qm}},c_{ij})$ decoheres.  For this purpose it
suffices to consider the non-negligible branch wave functions and examine
their overlap:
\begin{subequations}
\label{eq:Dijc}    
\begin{eqnarray}
   D(c_{ij},c_{kl}) &=&  {\mathcal N}\ 
        \Psi^{\mathit{WKB}}_{kl} \bullet \Psi^{\mathit{WKB}}_{ij}
                \label{eq:Dija}  \\  
          &=& {\mathcal N}\ 
              \langle q'' \| C_{kl} | \Psi^{\mathit{WKB}} \rangle \bullet
              \langle q'' \| C_{ij} | \Psi^{\mathit{WKB}} \rangle     .             
                \label{eq:Dijb}   
\end{eqnarray}
\end{subequations}
The integrals in (\ref{eq:Dijb}) are to be evaluated on $\sigma_0$.  According
to (\ref{eq:semiclWKBBWFij}), $\Psi^{\mathit{WKB}}_{ij}(q'')$ is peaked around
the classical paths in the class $c_{ij}.$ As at least one pair of the initial
or final intervals $\Delta'$ or $\Delta''$ will be disjoint on $\sigma_0$ (see
Figure \ref{fig:Psi_ij}) 
-- and even if one of the initial or final intervals coincide, the momentum 
support will be disjoint -- 
we see that the overlap integral in (\ref{eq:Dijc}) will be small, and so we
have approximate decoherence of the coarse-graining
$(c_{\mathit{qm}},c_{ij}):$
\begin{equation}
    D(\mathit{qm},ij)  \approx  0  
           \label{eq:WKBdecoh}
\end{equation}
In other words, the WKB branch wave functions approximately decohere because 
they are approximately macroscopically distinct.  
A higher degree of decoherence would require correlations with additional
degrees of freedom.

We now estimate the relative probabilities $p_{ij} = D(c_{ij},c_{ij})$ of the
various members of the family of classical paths preferred by the
WKB initial condition.  The answer will be a precise version of the familiar 
heuristic rule of quantum cosmology according to which these probabilities 
are given by fluxes throuh a spacelike surface in minisuperspace.  (See 
\cite{HalThor01,HalThor02} for related results.)

The branch wave functions $\Psi^{\mathit{WKB}}_{ij}(q'')$ have support on
$\sigma_0$ essentially only in $\Delta_i'$ and $\Delta_j''$.  Recall that in
Section \ref{sec:exampleinitial} we chose our initial state
$\Psi^{\mathit{WKB}}(q')$ to be positive frequency in $\alpha'$, and that, as
noted at the end of Section \ref{sec:cgsemiBWFs}, $\Psi^{\mathit{WKB}}_{ij}(q'')$
is negative frequency at the ends of the paths on $\sigma_0$. 
$\Psi^{\mathit{WKB}}_{ij}(q'')$ is therefore positive frequency in $\Delta_i'$
and negative frequency in $\Delta_j''$.  Further, because the current
$J_A = -(i/2)\Psi^*{\kern-0.35em}\buildrel\leftrightarrow\over{\nabla}_{\kern-0.35em A}{\kern-0.35em} \Psi$
is conserved 
and $\Psi^{\mathit{WKB}}_{ij}(q'')$ vanishes on any spacelike surface
$\sigma_f$ far to the future of $T_{ij}$ ({\it i.e.\ }at very large values of
$\alpha$ -- see Figure \ref{fig:Psi_ij}), integrating ${\nabla}_{\kern-0.35em
A}J^A = 0$ over a volume bounded by $\sigma_0$ and $\sigma_f$ we see that
\begin{subequations}
\label{eq:zeroflux}    
\begin{eqnarray}
    \int\nolimits_{\sigma_0} \hspace{-6pt} d \Sigma^A\  J_A 
        & = & 
        \int\nolimits_{\Delta_i'} \hspace{-7pt} d \Sigma^A\  J_A  +
        \int\nolimits_{\Delta_j''} \hspace{-7pt} d \Sigma^A\  J_A  
           \label{eq:zerofluxa} \\ 
        & = & 0
           \label{eq:zerofluxb}
\end{eqnarray}
\end{subequations}
from which we learn that
\begin{equation}
    \Psi^{\mathit{WKB}+}_{ij} \circ \Psi^{\mathit{WKB}+}_{ij}  +
    \Psi^{\mathit{WKB}-}_{ij} \circ \Psi^{\mathit{WKB}-}_{ij}  = 0
           \label{eq:fluxcancel}
\end{equation}
on $\sigma_0$.  Thus
\begin{subequations}
\label{eq:pijJ}    
\begin{eqnarray}
   p_{ij} &=&  {\mathcal N}\  
              \Psi^{\mathit{WKB}}_{ij} \bullet \Psi^{\mathit{WKB}}_{ij}
             \label{eq:pijJa}  \\  
          &=& {\mathcal N}\ \left(
              \Psi^{\mathit{WKB}+}_{ij} \circ \Psi^{\mathit{WKB}+}_{ij}  -
              \Psi^{\mathit{WKB}-}_{ij} \circ \Psi^{\mathit{WKB}-}_{ij} \right) 
             \label{eq:pijJb}  \\                            
          &=& 2\, {\mathcal N}\ 
              \Psi^{\mathit{WKB}+}_{ij} \circ \Psi^{\mathit{WKB}+}_{ij}  
             \label{eq:pijJc}  \\                            
          &\approx& 2\, {\mathcal N}\, 
              i\int_{\Delta_i'} d \Sigma'^A\ \Psi^{\mathit{WKB*}}_{ij}(q')
                 \buildrel\leftrightarrow\over{\nabla}_{\kern-0.35em A} 
                           \Psi^{\mathit{WKB}}_{ij}(q'). 
             \label{eq:pijJd}                                           
\end{eqnarray}
\end{subequations}
In (\ref{eq:pijJd}) the most obvious vestige of the effect of the class
operator $C_{ij}$ is the restriction, via (\ref{eq:semiclWKBBWFij}), of the
domain of integration on $\sigma_0$ to $\Delta_i'$, though the conditions on
the final position $\Delta_j''$ (equivalently, initial momentum $\Gamma_j'$)
corresponding to the class $c_{ij}$ are still present through
(\ref{eq:semiclWKBBWFij}) -- {\it cf.\ }(\ref{eq:Psi_ijDeltai}) -- and the
connection (\ref{eq:WKBmom}).

Inserting the semiclassical form (\ref{eq:WKBinitial}), dropping gradients of
the slowly varying factor $A$, and evaluating the remaining integral at the
center $q_i' = (\alpha_0,\vec{q}_i')$ of the domain of integration $\Delta_i'$
on $\sigma_0$,
\begin{subequations}
\label{eq:pijWKB}    
\begin{eqnarray}
   p_{ij} &\approx &  2\, {\mathcal N}\,  
              i\int_{\Delta_i'} d \Sigma'^A\ |A|^2\, 2i\nabla_{\kern-0.35em A} W
             \label{eq:pijWKBa}  \\                         
          &\approx &     
              - \,{\mathcal N}\,  |A(q_i')|^2\, \nabla_{\kern-0.35em A}' W(q_i')\, 
                        d\Sigma'^A_i
             \label{eq:pijWKBb}                          
\end{eqnarray}
\end{subequations}
when the projection of $\nabla'_{\kern-0.2em A} W(q')$ into $\sigma_0$ is
in $\Gamma_j'$, and is 
zero otherwise.  %
(A factor of 4 has been absorbed into the normalization factor ${\mathcal N}$ 
in the second line.)  %
If $\sigma_0$ is chosen to be a surface of constant $\alpha =
\alpha_0$ and recalling again that the initial state was chosen to be
``positive frequency'' so that $\partial W/\partial\alpha' < 0$,
(\ref{eq:pijWKBb}) reduces to
\begin{equation}
   p_{ij} \approx  
    \,{\mathcal N}\,  |A(q_i')|^2\, 
        \left|\frac{\partial W}{\partial\alpha'}(q_i')\right|\, \Delta_i'
\label{eq:pijWKBsigma0}    
\end{equation}
so long as $\partial W/\partial\vec{q}'(q_i') \in \Gamma_j'$. 
The 
factor ${\mathcal N}$ can be calculated in this
approximation by the requirement that the probabilities sum to unity.

Further, it may be worth noting that in arriving at (\ref{eq:pijJd}) the only
properties of the branch wave functions we employed were that the wave
functions (i) vanish as $\alpha \rightarrow \infty$; (ii) are positive
frequency in $\Delta_i'$; and (iii) negative frequency in $\Delta_j''$.  These
properties derive essentially from the corresponding properties of the
propagator, which in turn follow from our boundary conditions on the paths
-- namely, that they begin and end on $\sigma_0$ with ``expanding'' initial
and ``contracting'' final conditions -- and did not depend on any specific
property of the WKB form of the initial state.  Thus, for this class of
coarse-grainings we can write more generally
\begin{subequations}
\label{eq:pijJ+}    
\begin{eqnarray}
   p_{ij} &\approx&  -4\, {\mathcal N}\  
              \int\nolimits_{\Delta_i'} \hspace{-7pt} d \Sigma^A\  J^+_{ijA}
             \label{eq:pijJ+a}  \\  
          &\approx& - \, {\mathcal N}\ 
              J^+_{ijA}(q_i')\,  d \Sigma'^A_i
             \label{eq:pijJ+b}                              
\end{eqnarray}
\end{subequations}
on $\sigma_0$, where $J^+_{ijA}$ is the current constructed from the positive
frequency part of the branch wave function $\Psi_{ij}$ for {\it any}
initial state. %
(A factor of 4 has again been absorbed into ${\mathcal N}$ in the second 
line.)

We have thus recovered a version appropriate to the present formalism of the
familiar \cite{lesH,Hal91,Vil88} ``$J\cdot d\Sigma$'' rule of quantum
cosmology according to which $J_A d\Sigma^A$
($\approx |A|^2 \nabla_{\kern-0.35em A} W  d\Sigma^A$ for WKB initial states)  
gives the relative probabilities of the classical trajectories passing through
elements $d\Sigma^A$ of spacelike surfaces in minusperspace.  In our
formalism, however, that rule need only be applied on the initial surface
$\sigma_0$, and can be applied to initial states more general than just those
of WKB form.

\section{Conclusion}
\label{sec:conclusion}

We have constructed a sum-over-histories generalized quantum theory for 
the Bianchi IX cosmologies.  This fully four-dimensional formalism allows
predictions to be made concerning sets of (reparametrization invariant)
alternative histories of such a universe.  In particular, the predictions of
the theory are not restricted to those defined only at a moment of ``time", as
is typical of canonical theories, and hence are not restricted to alternatives
defined only by observables on superspace.  Many questions of physical
interest are not expressible in terms of alternatives ``at one moment of
time"; the generality of the alternatives about which it is possible to make
predictions, and the naturalness with which the formalism accommodates them,
are notable conceptual advantages of this quantization scheme.

Partitions of histories of physical interest include partitions according to
whether or not the universe becomes singular (by, for instance, the standard
discussed in Section \ref{sec:IXLambdaeq0}); partitions by ranges of values of
the volume and degree of anisotropy $\beta_{\pm}$ the universe has achieved a
given proper time after the initial condition; and partitions into classes of
those histories which are close to some classical cosmological trajectory, and
those which are not.  Approximate classicality, in the sense of obeying the
classical equations of motion, is exemplary of a prediction concerning an
alternative that is defined {\it over time}, not just at a single moment.  For
this last set of alternatives, we showed in Section \ref{sec:semiclassical}\
that, for indifferent final conditions and particular choices of initial
condition, the generalized quantum theory predicts with a probability near one
that the universe behaves semiclassically, approximately following the
classical trajector(ies) preferred by the initial condition.  This is a
sensible prediction, and stands in contrast to the less intuitive predictions
of some other quantizations of Bianchi IX (as mentioned at the
beginning of Section \ref{sec:semiclassical}.)  %

There is much that remains to be be done with this formulation of a quantum
mechanics for cosmology.  The most pressing issue concerns the development of
techniques for more careful evaluation of the class operators
(\ref{eq:threefourteen}) for various choices of coarse-graining, and a
correspondingly improved understanding of %
the limitations of the approximations employed in the present work.
Application of the general formalism described here to a detailed analysis of
specific examples of analytic or numerical classical solutions will then be of
interest.

Other issues include the extension of the definition of final boundary
conditions which are ``indifferent'' to the paths to the case where the
universe expands forever, thereby enabling consideration of ever-expanding
models and the calculation of the probability that a Bianchi IX universe with
cosmological constant will ``tunnel through'' to a state of eternal expansion. %
(Once the technical issues are sorted out, defining the decoherence functional
through (\ref{eq:Dindiff}) using the Rieffel induced innner product is 
likely to be a constructive approach to this problem.) %
Predictions relevant to modern day cosmological observations and questions of
environmental decoherence will require the inclusion of inhomogeneous modes of
matter fields and high frequency (gravity wave or graviton) modes of the
gravitational field.

In all cases, the {\it rules} for performing the required calculations are
clearly defined: specify the boundary conditions; set out the partition
defining the physical alternatives it is desired to study; compute the branch
wave functions (\ref{eq:branchwavefns}) corresponding to those partitions and
initial conditions, and employ the decoherence functional (\ref{eq:Dindiff})
to determine whether this set of alternatives is consistent.  If it is, the
diagonal elements of the decoherence functional inform us of the probabilities
for those alternatives according to (\ref{eq:threeone}).

Finally, it is perhaps worth emphasizing that the essence of the predictive
framework we have here described should be straightforward to generalize to
quantum-gravitational technologies -- string theory or loop quantum gravity,
for example -- other than the particular path-integral formulation we have
offered.  (The essential technical step lies in the definition 
of the restricted propagators (\ref{eq:threefour}).)

\begin{acknowledgments}

We would like to thank Don Marolf and John Whelan for useful conversations,
and Thea Howard for creating the figures.  This work was supported in part by
NSF Grants PHY-90-08502, PHY-95-07065, and PHY-02-44764.  D.C. was supported
in part also by the Canadian Institute for Theoretical Astrophysics (CITA) and
the Natural Sciences and Engineering Research Council of Canada (NSERC).

\end{acknowledgments}

\appendix* 

\section{Class Operators Satisfy Constraints}
\renewcommand{\theequation}{A.\arabic{equation}}

This appendix is devoted to demonstrating that the class operator matrix
element $\langle q^{\prime\prime} \| C_h \| q^\prime\rangle$ defined by
(\ref{eq:threefour}) satisfy an operator form of the constraints.  More
concretely, we show that, for the kind of coarse-grainings defining
classicality discussed in Section \ref{sec:semiclassical},
\begin{equation}
H_{q^{\prime\prime}} \left\langle q^{\prime\prime} \left\|C_h\right\|
q^\prime\right\rangle  = 0 
\label{eq:aone}
\end{equation}
when $q^{\prime\prime}$ is outside the region of configuration space where the
paths are limited by the coarse graining 
and is not equal to $q'$.  %
We consider a coarse graining where the paths to the future of some
$q^0=\alpha^*$ are constrained to lie in a tube $T$ about a particular
classical path.  We denote the class operator corresponding to this particular
coarse graining by $C_T$.  We will show that \eqref{eq:aone} holds for $\alpha
< \alpha^*$.  That can be used to show that the product (3.8) is independent
of the choice of initial hypersurface $\sigma^\prime$ and final hypersurface
$\sigma^{\prime\prime}$ provided they are both in the region $\alpha <
\alpha^*$.  There are other coarse grainings for which \eqref{eq:aone} holds
(see, {\it e.g.\ }\cite[Section VII]{lesH}.)  We focus on this one because it
is relevant for classicality.

It is convenient to work in $\dot N=0$ ``gauge'' of (\ref{eq:threethirteen}).   
The class operator matrix elements can then be written in the form  
(\ref{eq:threesixteen}) 
with
$\langle q^{\prime\prime} N\| C_T \|q^{\prime} 0\rangle$ given by 
\begin{equation}
\left\langle q^{\prime\prime} N\| C_T \|q^{\prime} 0\right\rangle = 
\int \delta q_{q'}^{q''}\, e_T\left[q^A (s)\right] 
 \exp \left(i\int\limits^N_0 ds\, 
  \left[\frac{1}{4}\ G_{AB}\ \frac{dq^A}{ds}\ \frac{dq^B}{ds} - V\right]\right).
\label{eq:atwo}
\end{equation}
Here the integral is over all paths connecting $q^\prime$ to
$q^{\prime\prime}$ and $e_T [q^A (s)]$ is the characteristic functional for
the tube: This is 1 for $\alpha< \alpha^*$ (paths unrestricted), 1 if the path
$q^A(s)$ lies in the tube for $\alpha > \alpha^*$, and 0 if any part lies
outside it for $\alpha >\alpha^*$.  \eqref{eq:atwo} is just
(\ref{eq:threeseventeenb}) written in terms of this characteristic functional.

We next show that $\langle q^{\prime\prime} N\| C_T \|q^\prime 0\rangle$
obeys the ``Schr\"odinger equation''
\begin{equation}
i\ \frac{\partial}{\partial N}\ \left\langle q^{\prime\prime} N\| C_T
\|q^\prime 0\right\rangle = H_{q^{\prime\prime}}  \left\langle
q^{\prime\prime} N\| C_T \|q^\prime 0\right\rangle.
\label{eq:athree}
\end{equation}
The demonstration is the same as Feynman's original demonstration that the
path integral defining the propagator in non-relativistic quantum
mechanics satisfies the Schr\"odinger equation \cite{Fey48}.  
The only novelty is the
characteristic functional $e_T[q^A(s)]$ in \eqref{eq:atwo}. We define the path
integral in \eqref{eq:atwo} as the limit of integrals over paths skeletonized
on $J+1$ uniformly spaced slices of $s$ as $J$ tends to infinity.
Specifically, let $s_0=0$, $s_1, s_2 \cdots$, $s_J=N$ be slices spaced by
$\epsilon = N/J$. Let $q^A_k$ be the value of $q^A$ on slice $k$. Then,
\begin{eqnarray}
\lefteqn{
\left\langle q^{\prime\prime} N\| C_T \|q^\prime 0\right\rangle =
  \lim\limits_{J\to\infty} \int \prod\limits^{J-1}_{k=1}
  \left(\prod\limits^3_{A=0} dq^A_k\right)\ 
  \left(\frac{-1}{4\pi i\epsilon}\right)^2
  \ e_T\left(q^A_k\right)
} \hspace{1.0cm}   \nonumber \\ 
  &  &
\times \exp \left\{i \sum\limits^J_{k=1} \epsilon\ \left[\frac{1}{4}\ G_{AB} 
\left(\frac{q^A_k - q^A_{k-1}}{\epsilon}\right)\ 
\left(\frac{q^B_k - q^B_{k-1}}{\epsilon}\right) - 
      V\left(q^A_k\right)\right]\right\}.
\label{eq:afour}
\end{eqnarray}
Here, $e_T(q^A)$ is the characteristic {\it function} for the tube equal to 1
inside the tube and zero outside for $\alpha > \alpha^*$ and equal to 1 for
$\alpha<\alpha^*$ (unrestricted).  The product of these functions on different
slices make up the functional $e_T[q^A(s)]$ in \eqref{eq:atwo}.  The measure
factor $(-1/4\pi i\epsilon)^2$  
can be deduced from the Liouville measure on phase-space
\cite{lesH,HarKuc86} 
recalling that $G_{AB} = {\rm diag}\ (-1, 1, 1, 1)$.

The successive integrations in \eqref{eq:afour} evolve $\langle
q^{\prime\prime} N\|C_T\| q^\prime 0)$ foward in $N$.  Thus,
\begin{eqnarray}
\lefteqn{
\left\langle q^{\prime\prime} N\| C_T \|q^\prime 0\right\rangle =
\lim\limits_{\epsilon\to 0} \int \prod\limits^3_{A=0} dq^A_{J-1} 
\left(\frac{-1}{4\pi i\epsilon}\right)^{\frac{1}{2}} e_T \left(q^A_J\right)
} \hspace{1.0cm}   \nonumber \\ 
  &  &
\times \exp \left\{i\epsilon\left[\frac{1}{4}\ G_{AB} 
\left(\frac{q^{\prime\prime A}-q^A_{J-1}}{\epsilon}\right)\ 
\left(\frac{q^{\prime\prime B} - q^B_{J-1}}{\epsilon} \right) - 
  V \left(q^A_J\right)\right]\right\}   \nonumber \\ 
  &  &   
  \times \left\langle q_{J-1} N - \epsilon\|C_T\|q^\prime 0\right\rangle.
\label{eq:afive}
\end{eqnarray}
Now restrict attention to $\alpha^{\prime\prime} < \alpha^*$ where
$e_T(q^A_J)=1$. The integration over $q^A_{J-1}$ is restricted by the
factor $e_T (q^A_{J-1})$ in $\langle q_{J-1} N-\epsilon \|C_T\|q^\prime 
0\rangle$. As $\epsilon \to 0$, an increasingly narrow range of $q^A_{J-1}$
near $q^{\prime\prime A}$ provides the only significant contribution to the
integral. The result is therefore no different from the integral without
any restriction in this limit for $\alpha^{\prime\prime} < \alpha^*$.
As in non-relativistic quantum mechanics \eqref{eq:afive}
implies the ``Schr\"odinger equation'' \eqref{eq:athree}.

Integrating \eqref{eq:athree} from $N=0$ to $N=\infty$ gives
\begin{equation}
H_{q^{\prime\prime}} \left\langle q^{\prime\prime} \|C_T \|q^\prime
\right\rangle = i\left[ \left\langle q^{\prime\prime} \infty \|C_T\| q^\prime 
0\right\rangle - \left\langle q^{\prime\prime} 0\|C_T \|  q^\prime
0\right\rangle \right]
\label{eq:asix}
\end{equation}
for $\alpha^{\prime\prime} <\alpha^*$. The first term on the right-hand 
side vanishes because what was a localized ``wave packet'' at $N=0$ has
spread over all of configuration space by $N=\infty$. The second term
vanishes if $q^{\prime\prime A} \not= q^{\prime A}$ for all $A$. That is
because for small $N=\epsilon$
\begin{equation}
\left\langle q^{\prime\prime} \epsilon\|C_T\| q^\prime 0\right\rangle =
\left(\frac{-1}{4\pi i\epsilon}\right)^2 \exp \left[i\epsilon\ \frac{1}{4}G_{AB}
\left(\frac{q^{\prime\prime A} - q^{\prime A}}{\epsilon}\right)
\ \left(\frac{q^{\prime\prime B} - q^{\prime B}}{\epsilon}\right)\right]
\label{eq:aseven}
\end{equation}
from \eqref{eq:afour} when $\alpha^{\prime\prime}< \alpha^*$.  But this is a
representation of a $\delta$-function as $\epsilon\to 0$, and we demonstrate
\eqref{eq:aone} for $\alpha^{\prime\prime} < \alpha^*$.  The right hand side
of \eqref{eq:aone} vanishes when $q' \neq q''$, and the class operators thus
in this sense satisfy an operator form of the constraints.  (More precisely,
we see that these class operators are Green functions for the Wheeler-DeWitt
Hamiltonian.
For more on the question of when class operators constructed according to 
(\ref{eq:threefourteen}) satisfy constraints, see 
\cite{Hal88,HalThor01,HH91,HM97,HalThor02,Hal04}.)


\def\acta{Acta Physica}
\def\advphys{Adv.\ Phys.\ }
\def\amjphys{Am.\ J.\ Phys.\ }
\def\annmath{Ann.\ Math.\ }
\def\annphys{Ann.\ Phys.\ }
\def\annrevastron{Ann.\ Rev.\ Astron.\ }
\def\apj{Ap.\ J.\, }
\def\apjlett{Ap.\ J.\ Lett.\ }
\def\astronj{Astronom.\ J.\ }
\def\astrophyslett{Astrophys.\ Lett.\ }
\def\canjmath{Canad.\ J.\ Math.\ }
\def\cqg{Class.\ Quant. Grav.\ }
\def\classquantgrav{Class.\ Quant. Grav.\ }
\def\cmp{Commun.\ Math.\ Phys.\ }
\def\foundphys{Found.\ Phys.\ }
\def\grg{Gen.\ Rel.\ Grav.\ }
\def\hadronic{Hadronic\ J.\ }
\def\intjmodphys{Int.\ J.\ Mod.\ Phys.\ }
\def\intjtheorphys{Int.\ J.\ Theor.\ Phys.\ }
\def\jlonmathsoc{J.\ London Math.\ Soc.\ }
\def\jmathmech{J.\ Math.\ Mech.\ }
\def\jmp{J.\ Math.\ Phys.\ }
\def\jphys{J.\ Phys.\ }
\def\jstatphys{J.\ Stat.\ Phys.\ }
\def\jsp{J.\ Stat.\ Phys.\ }
\def\camphilsoc{Math.\ Proc.\ Cam.\ Phil.\ Soc.\ }
\def\mnras{Mon.\ Not.\ R.\ Astron.\ Soc.\ }
\def\np{Nuc.\ Phys.\ }
\def\nucphys{Nuc.\ Phys.\ }
\def\nc{Nuovo Cim.\ }
\def\pr{Phys.\ Rev.\ }
\def\prd{Phys.\ Rev.\ }
\def\physrev{Phys.\ Rev.\ }
\def\prl{Phys.\ Rev.\ Lett.\ }
\def\pl{Phys.\ Lett.\ }
\def\physlett{Phys.\ Lett.\ }
\def\physrep{Phys.\ Rep.\ }
\def\pt{Physics Today}
\def\proysoclon{Proc.\ Roy.\ Soc.\ London}
\def\proysoced{Proc.\ Roy.\ Soc.\ Edinburgh}
\def\ptp{Prog.\ Theor.\ Phys.\ }
\def\progtheorphys{Prog.\ Theor.\ Phys.\ }
\def\qjras{Q.\ J.\ R.\ Astron.\ Soc.\ }
\def\rpg{Rep.\ Prog.\ Phys.\ }
\def\repprogphys{Rep.\ Prog.\ Phys.\ }
\def\rmp{Rev.\ Mod.\ Phys.\ }
\def\revmodphys{Rev.\ Mod.\ Phys.\ }


\def\APO{Academic Press, Orlando}
\def\Academic{Academic Press, Orlando}
\def\APNY{Academic Press, New York}
\def\AW{Addison-Wesley, Reading}
\def\Addison{Addison-Wesley, Reading}
\def\Birkhauser{Birkhauser, Boston}
\def\CUP{Cambridge University Press, Cambridge}
\def\Clar{Clarendon Press, Oxford}
\def\Clarendon{Clarendon Press, Oxford}
\def\Freeman{Freeman, San Francisco}
\def\Gordon{Gordon and Breach, New York}
\def\IOP{Institute of Physics, Bristol}
\def\KNY{Kluwer, New York}
\def\Kluwer{Kluwer, New York}
\def\KD{Kluwer, Dordrecht}
\def\NYAS{New York Academy of Science, New York}
\def\NH{North Holland, Amsterdam}
\def\OUP{Oxford University Press, Oxford}
\def\Oxford{Oxford University Press, Oxford}
\def\Japan{Physical Society of Japan, Tokyo}
\def\Plenum{Plenum, New York}
\def\Pol{Polish Scientific Publishers, Warsaw}
\def\PUP{Princeton University Press, Princeton}
\def\Reidel{Reidel, Boston}
\def\Springer{Springer-Verlag, Berlin}
\def\UCP{University of Chicago Press, Chicago}
\def\Chicago{University of Chicago Press, Chicago}
\def\Wiley{Wiley, New York}
\def\WS{World Scientific, Singapore}
\def\World{World Scientific, Singapore}

%
%
%
%

\gdef\journal #1, #2, #3, #4          {%
,\ {\it #1}{\bf #2} #3 (#4).        }   

\gdef\onejournal #1, #2, #3, #4          {%
\ {\it #1}{\bf #2} #3 (#4).        }   


\gdef\ljournal #1, #2, #3, #4 {%
,\ {\it #1}{\bf #2} #3 (#4)}

\gdef\oneljournal #1, #2, #3, #4 {%
;\ {\it #1}{\bf #2}, #3 (#4)}

%
%

\gdef\book #1, #2, #3, #4     {%
#1, {\it #2} (#3, #4).       }

\gdef\lbook #1, #2, #3, #4     {%
#1, {\sl #2} (#3, #4)}

\gdef\edbook #1, #2, #3, #4     {%
{\it #2}, #1 (#3, #4).       }

\gdef\ledbook #1, #2, #3, #4     {%
{\it #2}, #1 (#3, #4)}

\def\qcbu{\ledbook {edited by S.~Coleman, J.~B.~Hartle, T.~Piran, and 
S.~Weinberg}, Quantum Cosmology and Baby Universes: Proceedings of the 1989 
Jerusalem Winter School for Theoretical Physics, \World, 1991 }

\def\julia{\ledbook {edited by B.~Julia and J.~Zinn-Justin},
Gravitation and Quantizations, \NH, 1995 }

\def\cepi{\ledbook {edited by W. Zurek},  {Complexity, Entropy, and the 
Physics of Information, SFI Studies in the Sciences of Complexity, 
Volume VIII}, \Addison, 1990 }

\def\newtech{\ledbook {edited by S.~Kobayashi, H.~Ezawa, M.~Murayama, 
and S.~Nomura}, Proceedings of the 3rd International Symposium on the
Foundations of Quantum Mechanics in the Light of New Technology, 
\Japan, 1990 }

\def\hpm{\ledbook {edited by J.~J.~Halliwell, J.~ Perez-Mercader, 
and W.~Zurek}, Proceedings of the NATO Workshop on the Physical Origins 
of Time Asymmetry, \CUP, 1994 }

\def\giulini{\lbook {E.~Joos, H.~D.~Zeh, C.~Kiefer, D.~Giulini, J.~Kupsch, and I.-O.~Stamatescu}, 
{Decoherence and the Appearance of a Classical World in Quantum Theory, Second Edition},  
\Springer, 2003 }

\def\qg2{\ledbook {edited by C.~J.~Isham, R.~Penrose, and 
D.~W.~Sciama}, Quantum Gravity 2, \Clarendon 1981 }

\def\weinberg{\book S.~Weinberg, Gravitation and Cosmology, \Wiley, 
1972}

\def\zeh{\book H.~D.~Zeh, The Physical Basis of the Direction of Time, Fourth Edition, 
\Springer, 2001}

\end{document}